\documentclass[aps,pra,onecolumn,showpacs,showkeys,floatfix,groupedaddress,longbibliography]{revtex4-2}

\usepackage{amsfonts,fancybox,pgf}
\usepackage{amssymb}
\usepackage{float}
\usepackage{footnote}
\usepackage{amsmath}
\usepackage{makeidx}
\usepackage[ansinew]{inputenc}
\usepackage[usenames,dvipsnames]{pstricks}
\usepackage{enumerate}                  
\usepackage{footnote}                   
\usepackage{microtype}                  
\usepackage{rotfloat}    
\usepackage{graphicx}
\usepackage{here}
\usepackage{epstopdf}
\usepackage{epsfig}
\usepackage{pst-grad} 
\usepackage{pst-plot} 
\usepackage[colorlinks,hyperindex]{hyperref}
\usepackage[english]{babel}
\usepackage[caption=false]{subfig}
\usepackage[top=3cm,bottom=3cm,left=3.2cm,right=3.2cm,headsep=13pt,a4paper]{geometry}
\usepackage[toc,page]{appendix}
\usepackage{chngcntr}

\hypersetup
		{
		colorlinks,%
		citecolor=blue,%
		linkcolor=black,%
		urlcolor=black,%
	}

\makeindex

\begin{document}

\title{\normalsize Confinement-Tunable Synthetic Gauge Fields and Floquet Topological Phenomena in a Driven Quantum Wire Qubit \\}

\author{S. L. Dongmo Tedo$^{b}$, O. C. Feulefack$^{a}$, J. E. Danga$^{b, c, d}$,  R.M. Keumo Tsiaze$^{a, c, e}$, F. Melong$^{f}$, C. Kenfack Sadem$^{a, b, c}$,  A. J. Fotue$^{b}$, M. N. Hounkonnou$^{a}$, L.C. Fai$^{b, c}$}

\address{$^{a}$ International Chair in Mathematical Physics and Applications, (ICMPA-UNESCO Chair), University of Abomey-Calavi, 072 P O Box 50, Cotonou, Republic of Benin.}
\address{$^{b}$ Condensed Matter and Nanomaterials, Department of Physics, Faculty of Science, University of Dschang,P.O. Box 67, Dschang, Cameroon.}
\address{$^{c}$ Quantum Materials and Computing Group - QMaCG, P.O. Box 70 Bambili, Northwest Region, Cameroon.}
\address{$^{d}$ Instituteur principal de l'enseignement g\'{e}n\'{e}ral, D\'{e}l\'{e}gation R\'{e}gionale de l'Ouest,  D\'{e}l\'{e}gation D\'{e}partementale de la Menoua, Arrondissement de Nkong-Ni, \'{E}cole Publique de Batsing'la Chefferie, Bafou, Cameroun.}
\address{$^{e}$ Laboratory of Mechanics, Materials and Structures, Faculty of Science, University of Yaound\'{e} I, P.O. Box 812, Yaound\'{e}, Cameroon.}
\address{$^{f}$ Mathematisches Institut der Universit$\ddot{a}$t M$\ddot{u}$nster Einsteinstr. 62, 48149 M$\ddot{u}$nster, Germany.}
\email{edmonddanga1@yahoo.com, keumoroger@gmail.com, hounkonnou@yahoo.fr, corneliusfai@gmail.com}

\begin{abstract}
Theoretical analysis indicates that a spin qubit in a parabolic quantum wire, when driven by a bichromatic field, generates a confinement-tunable non-Abelian synthetic gauge field that gives rise to novel Floquet topological phenomena. In the low-confinement regime, the system exhibits topological Landau-Zener (LZ) transitions, chiral Landau-Zener-St$\ddot{u}$ckelberg-Majorana (LZSM) interference, and non-Abelian geometric phases around diabolical points. These features enable holonomic quantum computation, as well as unconventional Floquet Bloch oscillations and fractal spectra. In contrast, in the strong confinement regime, narrow avoided crossings significantly suppress cross-coupling, thereby insulating the qubit from multi-photon LZSM interference. Quantitative Floquet-Lindblad analysis shows that the optimized gate times, determined by the biharmonic driving period and non-Abelian parameter-space loops, are sufficiently short to surpass primary environmental decoherence channels. When applied to high-mobility $\text{GaAs/AlGaAs}$ core-shell nanowires, this approach converts spatial drives into robust non-Abelian gauge fields, positioning quantum wire systems as a versatile and scalable platform for fault-tolerant quantum information processing.
\end{abstract}

\vspace{0.5pc}


%

\maketitle
%

\section{Introduction}\label{sec1}
\noindent

The development of fault-tolerant quantum information processing has stimulated extensive research into platforms that facilitate coherent control and protect quantum states from decoherence. Semiconductor quantum wires, characterized by high integrability and tunable electronic confinement, constitute a promising architecture for hosting spin qubits \cite{Loss, Hanson, Kloeffel, Auslaender, Tserkovnyak, Tserkovnyak2, Auslaender2, Steinberg, Okuda, Shendryk}. However, maintaining quantum coherence while achieving high-fidelity operations remains a significant challenge.

Floquet engineering, which involves controlling quantum systems through periodic driving, has become a prominent approach for generating non-equilibrium quantum phases with enhanced robustness \cite{Oka, Runder, Bukov}. Periodically driven (``Floquet") two-level systems now serve as a central platform for engineering effective Hamiltonians that are unattainable in static environments, enabling phenomena such as dynamical localization and synthetic gauge fields \cite{Bukov, Grifoni}.

Within this framework, LZSM interferometry is a central technique for probing and manipulating driven qubits  \cite{Shevchenko, Forster, Pedersen}. When the drive includes two independent quadrature components that form a biharmonic field, the resulting quasi-energy spectrum exhibits a complex multiphoton resonance structure. The corresponding LZSM interference pattern provides a direct spectroscopic probe of the underlying control landscape \cite{Shevchenko, Ivakhnenko, Grossman, Berry}. This interference phenomenon, initially analyzed in the weak-driving, single-crossing regime, is now recognized to support Coherent Destruction of Tunneling (CDT) at specific points where the effective coupling vanishes exactly \cite{Grossman}. Furthermore, the high-frequency and scaling properties of such driven systems are now well characterized within the broader context of Floquet engineering \cite{Berry}.

Since Berry's original work, it has been established that parameter-dependent degeneracies of a real Hamiltonian, known as ``diabolical points", are a generic feature of any two-parameter family rather than a fine-tuned coincidence \cite{Berry2, Berry3}. Adiabatic transport around such a point results in the accumulation of a quantized geometric phase \cite{Zanardi}. When a degenerate multiplet, rather than an isolated pair, is transported, this phase \cite{Lindner, Wilczek} becomes matrix-valued and non-Abelian \cite{Okuda}, forming the basis for holonomic quantum computation \cite{Lauhon}. The realization and control of such non-Abelian holonomies in a solid-state qubit would unify two typically distinct research areas: driven-qubit LZSM interferometry and topological or holonomic gate protocols, within a single, gate-voltage-tunable device.

Recent advances indicate that stronger electronic confinement can improve control over rapid quantum tunneling, thereby preserving spin states during LZSM interferometry transitions \cite{Danga, Danga2, Mkam, Danga3}. While monochromatic drives have been extensively investigated \cite{Oliver, Berns}, biharmonic fields provide a broader control landscape and greater phase sensitivity \cite{Satanin, Blattmann, Moessner, Weitenberg}. Prior studies have focused on minimizing noise effects \cite{You1, You2} and examining nonlinear qubit dynamics \cite{Sillanpaa, Wilson, Izmalkov, SGasparinetti, Berns2, Rudner}  to regulate topological photonic gap magnitudes across diverse field-induced energy spectra. Semiconductor nanowires represent a promising platform for these investigations. Gate-defined confinement in III-V heterostructures enables continuously tunable effective potentials \cite{Funk, NadjPerge}, and spin-orbit-coupled nanowire qubits driven by alternating current (AC) electric or magnetic fields \cite{Rui, Rui2} via electric-dipole spin resonance have demonstrated coherent, microwave-frequency control of effective two-level systems with millitesla-scale field amplitudes \cite{Forster, Pedersen}. Core-shell growth techniques that produce Majorana-relevant nanowire heterostructures \cite{Lutchyn, Nayak} also yield high-mobility, modulation-doped GaAs/AlGaAs core-shell nanowires with well-controlled conduction-band interfaces \cite{Funk, Wilczek}. As a result, a gate-tunable, biharmonically driven qubit in such a nanowire constitutes an experimentally viable, near-term objective rather than a purely theoretical construct. Nevertheless, the dynamics of qubits exposed to biharmonic fields, particularly in conjunction with dynamically tunable confinement, remain insufficiently characterized. Additionally, these approaches have been proposed to improve precision measurement in quantum metrology \cite{HZhou, Fiderer, MJiang} and to facilitate advances in holonomic quantum computing \cite{Pachos, Zanardi, Pachos2}.

In this work, we investigate a two-level system embedded in a one-dimensional, gate-tunable parabolic confinement, driven by a biharmonic microwave field characterized by frequency $\omega$, amplitudes $\alpha$ and $\beta$, and relative phase $\theta$. We demonstrate that the confinement frequency $\Omega$ influences the effective bias exclusively through the ratios $\alpha/\omega$ and $\beta/\omega$, enabling continuous tuning between a perturbative, single-resonance Floquet spectrum and a multiphoton, multi-crossing regime. Employing the exact Floquet formalism \cite{HuWu, Xueda, Scully} with the rotating-wave approximation, we calculate the quasi-energy gap \cite{Shirley} and the one-period LZSM transition probability across the ($\alpha/\omega$, $\beta/\omega$), ($\beta/\omega$, $\omega$), and ($\beta/\omega$, $\theta$) control planes. We identify the resulting diabolical points as genuine codimension-2 degeneracies of the underlying real Hamiltonian \cite{Berry} in the low-confinement regime, and construct a quantitative topological phase diagram N($\Omega$), which represents the number of such points accessible within a realistic drive-amplitude window. This formalizes the confinement-induced topological Landau-Zener transition \cite{Zener, Khomeriki}. Additionally, we define and evaluate the associated gate time, demonstrating that the diabolical points enabling non-Abelian holonomic control in the low-confinement regime coincide with the divergence of the naive dynamical gate time.

Unless the system is tuned precisely to resonance ($\gamma_1 = m\omega$), the detuning term $\delta$ dominates the extremely small effective gap in the high-confinement regime. When the phase $\theta$ or the amplitudes are adiabatically varied over a closed loop, the quantum state acquires only an Abelian U(1) geometric phase. The state vector accumulates a complex phase factor $\exp(i\gamma)\Big|_{\gamma = 1, 2}$, but its orientation in Hilbert space remains unchanged. The quasi-energies are given by $\pm\delta/2$, and the maximum probability amplitude for reaching the excited state approaches zero. Higher-order photon transitions ($m \geq 2$) are strongly suppressed, resulting in the qubit remaining predominantly in its initial state. The full construction is mapped onto a concrete, fabricable $\text{GaAs/AlGaAs}$ high-mobility core-shell nanowire device.

The structure of this paper is as follows. Section II introduces the model Hamiltonian in a two-level approximation, clarifies the control parameters, and analyzes the dynamics of topological qubit states under biharmonic driving and curved confinement. Section III examines the quasi-energies and transition probabilities, providing analytical results in the high- and low-confinement limits and numerical results under the rotating wave approximation (RWA). This spectral analysis informs the study of LZSM interferometry transitions under the specified driving protocol, identifies diabolical points, and enables the observation of novel Floquet topological phenomena \cite{Silaev}. Section IV presents a numerical investigation of the physical realization in a $\text{GaAs/AlGaAs}$ core-shell nanowire, evaluates gate times, and assesses the impact of decoherence. Section V discusses the results and offers concluding remarks.

\section{Model}\label{sec2}
\noindent

This study examines a spin qubit in a three-dimensional heterostructure magnetic quantum wire. The system is subjected to a parabolic confinement potential of defined strength and a biharmonic electromagnetic field. The presented Hamiltonian characterizes a longitudinally driven qubit, where the drive is aligned with the $\sigma_z$ axis and the static coupling is oriented along $\sigma_x$. The drive incorporates both single-photon (frequency $\omega$) and two-photon (frequency $2\omega$) harmonic components. The system is governed by the following time-dependent Hamiltonian \cite{Danga4}:
\begin{equation}
\mathcal{H}_\text{LZSM}(t) = \frac{-h(t)}{2}\hat{\sigma}_z - \frac{\Delta}{2}\hat{\sigma}_x
\end{equation}
where $\hat{\sigma}_{x,z}$ are Pauli matrices, $\Delta$ is the tunnelling matrix element, and the time-dependent bias is
\begin{equation}
h(t) = \gamma_1 + \gamma_2\sin(\omega{t}) - \gamma_3\cos(2\omega{t} + \theta).
\end{equation}
The coefficients $\gamma_i$ explicitly depend on the confinement $\Omega$ and the magnetic field amplitudes ($\alpha = \mu_BB_0$,  $\beta = \mu_BB$):
\begin{equation}
\gamma_1 = 1 + \frac{\alpha^2}{2\Omega^2} + \frac{\beta^2}{4\Omega^2}, \phantom{.} \gamma_2 =\frac{\alpha\beta}{\Omega} \phantom{.} \textrm{and} \phantom{.} \gamma_3 = \frac{\beta^2}{2\Omega}.
\end{equation}
We assume that $B_0$ is a tunable homogeneous magnetic field perpendicular to the plan of the wire and $B$ is the amplitude of the magnetic field. The direct coupling between spatial confinement and temporal drive parameters facilitates the novel effects described in this study. To analyse this periodically driven system, the Floquet formalism is employed \cite{Danga, Danga2}. Consider $|\psi(t)\rangle$ as defined by:
\begin{equation}
|\psi(t)\rangle = V_0(t)|\bar{\psi}(t)\rangle,
\end{equation}
which represent periodic Floquet states, where
\begin{equation}
V_0(t) = \exp\Big[\frac{-i}{2\hbar}\phi(t)\sigma_z\Big],
\end{equation}
a canonical transformation is subsequently applied. Since the time-dependent drive is purely longitudinal along $\sigma_z$, it is possible to transition into a rotating frame to eliminate the rapidly oscillating diagonal term. A unitary transformation is then implemented, as described in Eq. (5), where $\phi(t)$ denotes the integral of the purely time-dependent component of the $z$-field:
\begin{equation}
  \begin{split}
\phi(t) & =   \int_0^t \big[\gamma_1 - \gamma_2\sin(\omega{\tau}) - \gamma_3\cos(2\omega{\tau} + \theta)\big]d\tau \\
        &=   \gamma_1{t} + \frac{\gamma_2}{\hbar\omega}\cos(\omega{t}) - \frac{\gamma_3}{2\hbar\omega}\sin(2\omega{t} + \theta) + K(\gamma_2, \gamma_3).              
\end{split}
\end{equation}
The function $K(\gamma_2, \gamma_3) = \frac{\gamma_3}{2\hbar\omega}\sin\theta - \frac{\gamma_2}{\hbar\omega}$ represents a constant of integration. Transforming the transverse coupling into the interaction picture yields a new Hamiltonian in the corresponding rotating frame.
\begin{equation}
\bar{H}(t) = V_0^{\dag}H(t){V_0} - i\hbar{V_0^{\dag}}\frac{\partial{V_0}}{\partial{t}}.
\end{equation}
The Schr$\ddot{o}$dinger for the transformed wave-function $|\bar{\psi}(t)\rangle$ takes the form
\begin{equation}
i\hbar\frac{\partial}{\partial{t}}|\bar{\psi}(t)\rangle = \Big(V_0^{\dag}H(t){V_0} - i{V_0^{\dag}}\frac{\partial{V_0}}{\partial{t}}\Big)|\bar{\psi}(t)\rangle = \bar{H}(t)|\bar{\psi}(t)\rangle.
\end{equation}
A formal solution to the Schr$\ddot{o}$dinger equation (Eq. 8) is expressed as $|\psi(t)\rangle = U(t, t_0)|\psi(t_0)\rangle$.
\begin{equation}
U(t, t_0) = \hat{\zeta}\exp\Big(-\frac{i}{\hbar}\int^t_{t_0}H(\tau)d\tau \Big).
\end{equation}
$\hat{\zeta}$ represents the time-ordering operator. The time evaluation over one period is described by the operator
\begin{equation}
U(T) = U(t + T, t) = \hat{\zeta}\exp\Big(-\frac{i}{\hbar}\int^{t + T}_{t}H(\tau)d\tau \Big).
\end{equation}
This operator is called the Floquet operator \cite{Grifoni, Shirley}. The eigenvalues of the Floquet operator are given by
\begin{equation}
U(T)|\Phi_k(t)\rangle = \exp\Big(-i\frac{E_k{T}}{\hbar}\Big)|\Phi_k(t)\rangle, \phantom{......} |\Phi_k(t+T)\rangle = |\Phi_k(t)\rangle.
\end{equation}
In this system, the parameters $E_k$ are called quasi-energies and will be defined in subsequent sections. Following the computational procedures detailed in {\color{blue} appendix A}, the transformed Hamiltonian takes the form:
\begin{equation}
\bar{H}(t) = -\frac{\Delta}{2}\big[\hat{\sigma}_x\cos{\phi(t)} + \hat{\sigma}_y\sin{\phi(t)}\big],
\end{equation}
which can equivalently be written in matrix form as 
\begin{equation}
\bar{H}(t) = -\frac{\Delta}{2}\begin{pmatrix}
0&e^{-i\phi(t)}\\e^{+i\phi(t)}&0
\end{pmatrix}.
\end{equation}
The result in Eq. (13) is significant: the canonical transformation $V_0(t)$ fully incorporates the dynamic bias $h(t)$ into the phase $\phi(t)$. The resulting Hamiltonian retains only the tunneling coupling $\Delta$, which is modulated by the oscillating phase. From this Hamiltonian, the Fourier-Jacob-Anger expansion is applied to derive the Floquet matrix elements and the transition probability. The transverse component can be rewritten using the raising and lowering operators as follows:
\begin{equation}
\bar{H}(t) = -\frac{\Delta}{2}\big(e^{-i\phi(t)}\sigma_{+} + e^{i\phi(t)}\sigma_{-}\big).
\end{equation}

\subsection{Rotating Wave Approximation (RWA) Conditions}
\noindent

Equations 13 and 14 show that the expressions $\exp(+i\phi(t))$ and $\exp(-i\phi(t))$ are present in the Hamiltonian. Factoring $\exp(\pm i\phi(t))$ by isolating the individual contributions yields four distinct exponential terms:
\begin{equation}
\exp\big[\pm{i}\phi(t)\big]  = \exp{\pm{i}\gamma_1{t}}\cdot\exp{\pm{i}K(\gamma_2, \gamma_3) }\cdot\exp\Big(\pm{i}\frac{\gamma_2}{\hbar\omega}\cos(\omega{t})\Big)\cdot\exp\Big(\pm{i}\frac{\gamma_3}{2\hbar\omega}\sin(2\omega{t} + \theta)\Big).
\end{equation}
The Jacobi-Anger expansions are employed to address the oscillatory terms,
\begin{equation}
\exp{\Big(\pm{i}\frac{\gamma_2}{\hbar\omega}\cos(\omega{t})\Big)} = \sum_{n = -\infty}^{+\infty} i^n\cdot{J}_n\Big(\frac{\gamma_2}{\hbar\omega}\Big)\cdot{e^{\pm{i}{n}\omega{t}}},
\end{equation}
\begin{equation}
\exp{\Big(\pm{i}\frac{\gamma_3}{2\hbar\omega}\sin(2\omega{t} + \theta)\Big)} = \sum_{m = -\infty}^{+\infty} {J}_m\Big(\frac{\gamma_3}{2\hbar\omega}\Big)\cdot{e^{\pm{i}{m}\omega{t}}}\cdot{e^{\pm{i}{m}\theta}},
\end{equation}
where ${J}_{i=m, n}$ denotes the Bessel function of the first kind. Using the following expression to represent the phase factor associated with the \(n^{\text{th}}\) and \(m^{\text{th}}\) harmonics of the driving field: 
\begin{equation}
\Theta_{n, m} = i^ne^{\pm{i}m\theta},
\end{equation}
the final expanded form is derived from the computations presented in {\color{blue} appendix B}.
\begin{equation}
\exp\big[\pm{i}\phi(t)\big] = \exp{\pm{i}K(\gamma_2, \gamma_3)}
\cdot\sum_{m, n = -\infty}^{+\infty}\Theta_{n, m}{{J}_n\Big(\frac{\gamma_2}{\hbar\omega}\Big)}\cdot{{J}_m\Big(\frac{\gamma_3}{2\hbar\omega}\Big)}\cdot{e^{\pm{i}({\gamma_1 + n\omega + 2m\omega)t}}}.
\end{equation}
The mathematical expression denoted by $\Theta_{n,m}$ varies considerably across different mathematical and physical contexts, as this notation is not universally standardized. After time-averaging, only terms for which the time-exponential factors cancel, specifically when $\gamma_1 + (n + 2m)\omega \approx 0$, persist. This leads to the condition $n + 2m = -N$. When the RWA is applied near an N-photon resonance, where the static splitting satisfies $\gamma_1 \approx N\omega$, rapidly oscillating terms are eliminated. Consequently, the effective renormalized coupling $\Delta_{\text{eff}}$, which determines the qubit's dynamics in the rotating frame, is weighted by this complex phase factor:
\begin{equation}
\Delta_{\text{eff}}^{n, m}(\Omega, \theta) = \Delta\sum_{n, m = -\infty}^{+\infty}\Theta_{n, m}{{J}_n\Big(\frac{\gamma_2}{\hbar\omega}\Big)}\cdot{{J}_m\Big(\frac{\gamma_3}{2\hbar\omega}\Big)}
\end{equation}
This phase factor provides a crucial geometric control parameter. By varying the phase $\theta$ of the $2\omega$ drive, it is possible to manipulate the complex argument of $\Delta_{\text{eff}}^{n, m}$, thereby steering the qubit's evolution trajectory across the Bloch sphere to implement the holonomic gates described below. 

\begin{figure}
\begin{center}
	\includegraphics[width=14cm]{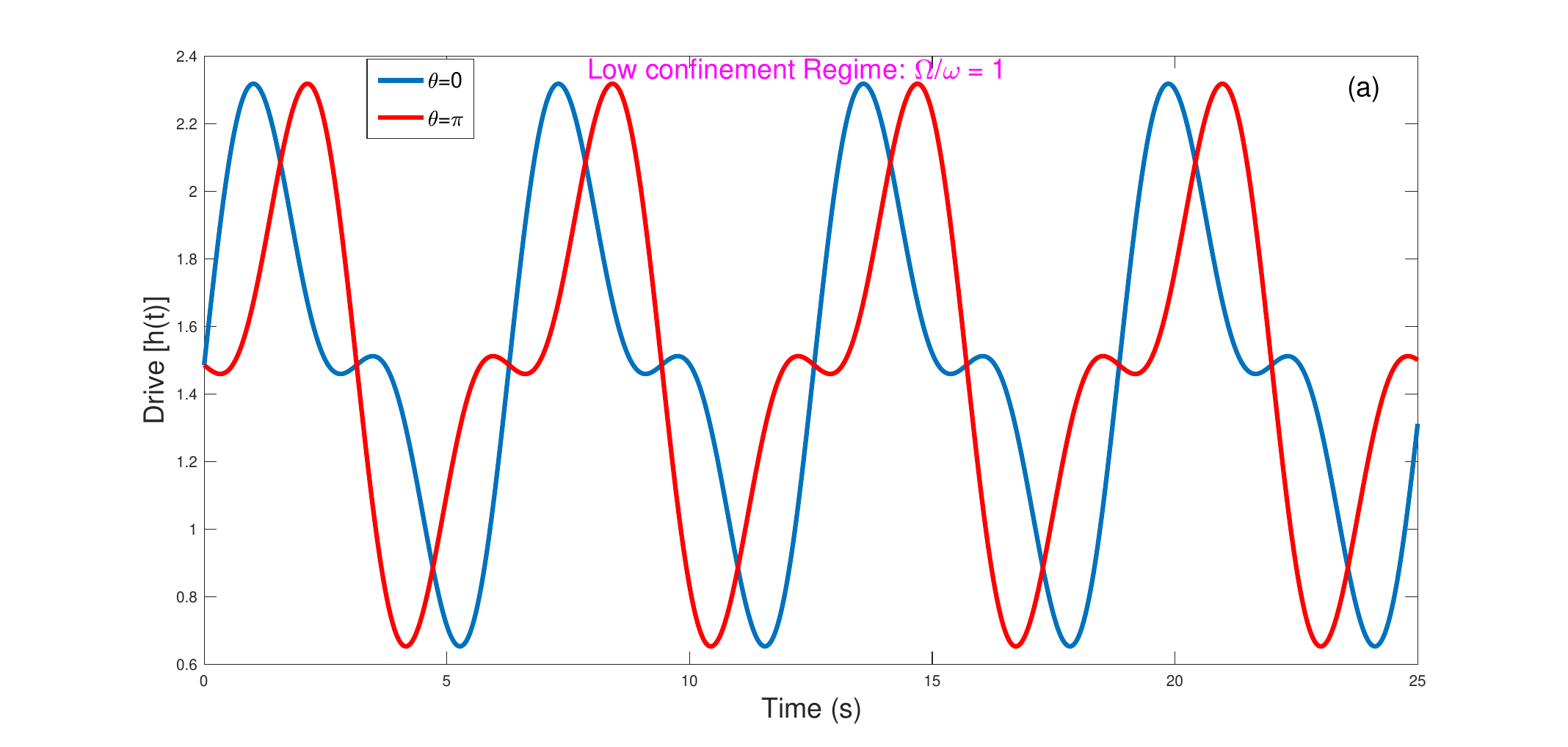}
	\includegraphics[width=14cm]{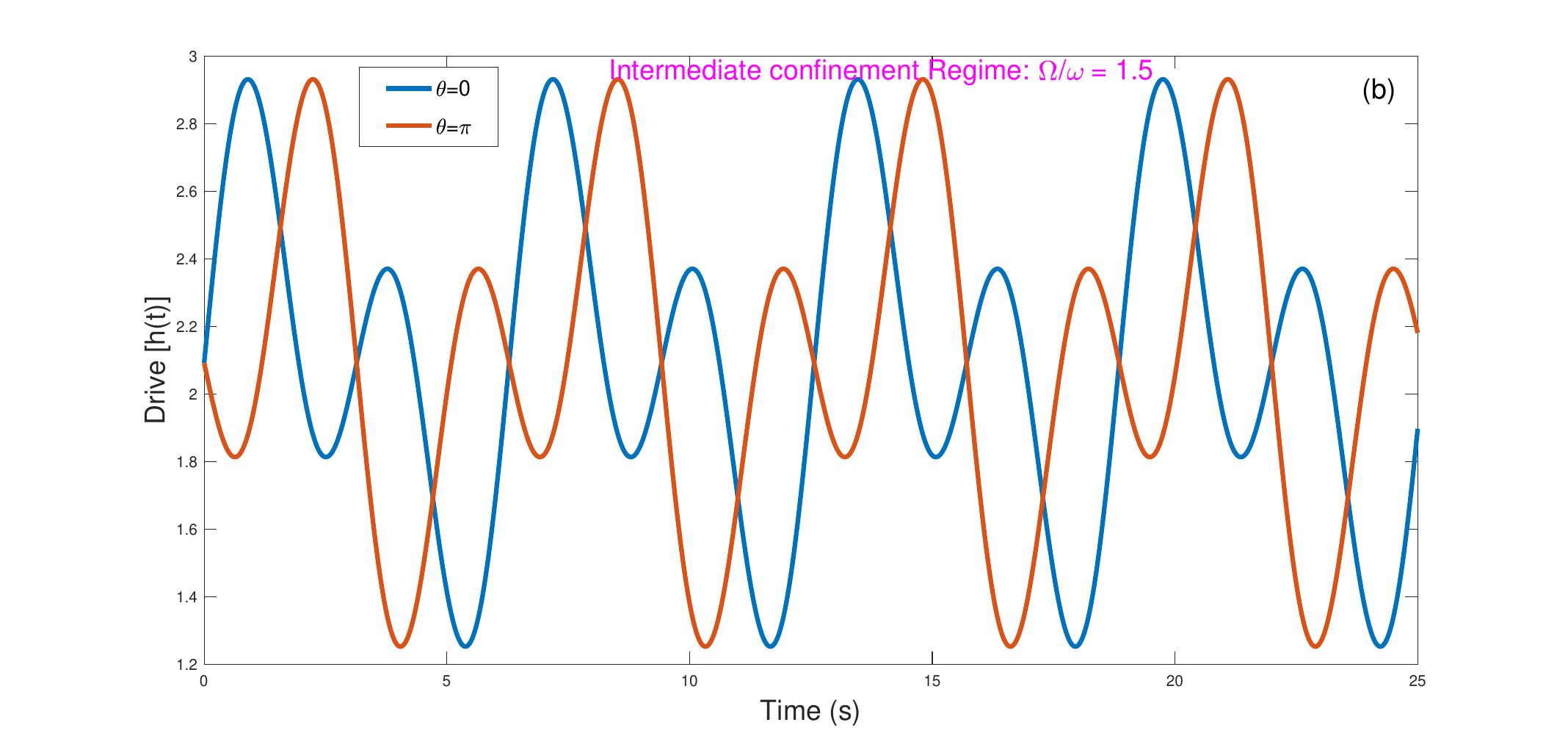}
	\includegraphics[width=14cm]{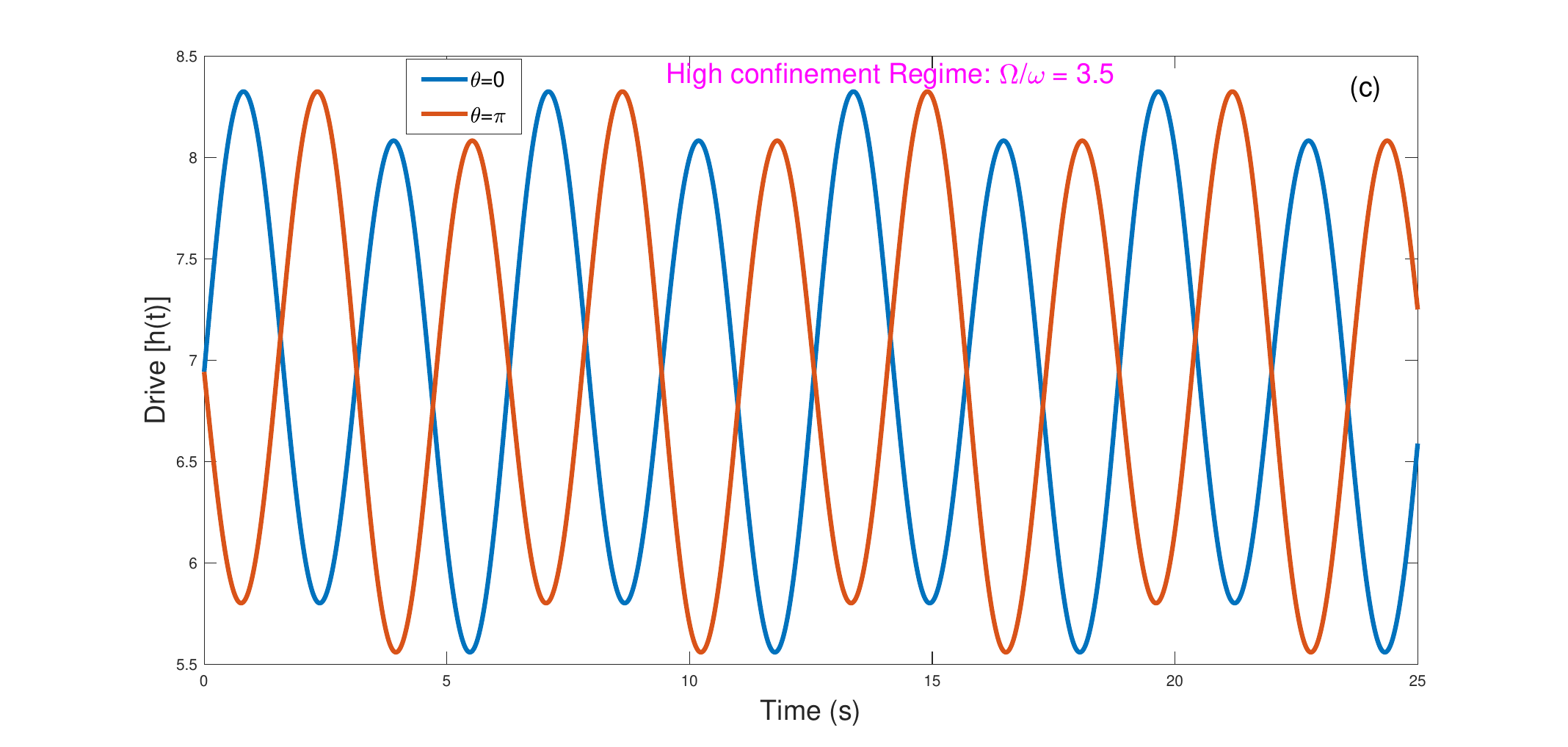}
\caption{(Color online) Drive waveforms are shown for different values of curved confinement: (a) $\Omega/\omega$ = 1, (b) $\Omega/\omega$ = 1.5, and (c) $\Omega/\omega$ = 3.5, with the drive asymmetry parameter fitted at $\theta = 0$ and $\theta = \pi$. The waveform becomes asymmetric in time when $\theta \neq 0$. As the curved confinement increases, successive phases of the drive signal lead to a buildup of excited-state population. The system parameters are $\alpha/\omega$ = 0.5 and $\beta/\omega$ = 1.2.}
\end{center}
\end{figure}

The quasi-energies $E_{1, 2}^{n, m}$ are the eigenvalues of the effective Floquet Hamiltonian which dictates the energy levels of the system dressed by the driving field:
\begin{equation}
E_{1, 2}^{n, m}(\gamma_2, \gamma_3, \theta) = \pm\frac{1}{2}\sqrt{\delta^2 + |\Delta_{\text{eff}}^{n, m}|^2}
\end{equation}
The magnitude of this effective coupling, $|\Delta_{\text{eff}}^{n, m}|$, serves as the renormalized Rabi frequency, while $\delta = \gamma_1 - N\omega$ denotes the detuning. Each term in the double sum of Eq. (19) has a well physical meaning as summarized in the table 1 below:

\begin{table}[htbp]
  \centering
  \caption{Physical interpretation of the term appearing in the Jacobi-Anger expansion}
\begin{tabular}{|c|c|c|}
\hline
  \textbf{N}$^o$  & \textbf{Terms}                                 & \textbf{Physical role}                                      \\
\hline  
  1               & $\phantom{.....}$  ${J}_n\Big(\frac{\gamma_2}{\hbar\omega}\Big)$ $ \phantom{.....}$  & Coupling in a fundamental mode $\omega$ (field $\gamma_2$)  \\
\hline
  2               & $\phantom{.....}$ ${J}_m\Big(\frac{\gamma_3}{2\hbar\omega}\Big)$ $\phantom{.....}$  & Coupling in a fundamental mode $2\omega$ (field $\gamma_3$) \\
\hline  
  3               & $\phantom{.....}$  ($n + 2m$)  $\phantom{.....}$                                    & Net number of photons exchanged                              \\
 \hline
\end{tabular}
\end{table}

Figure 1 illustrates the driving waveform $h(t)$ across different confinement strengths. At low confinement ($\Omega/\omega = 1$), the waveform exhibits significant asymmetry for $\theta \neq 0$. As $\Omega/\omega$ increases ($\Omega/\omega = 3.5$), symmetry in the waveform is restored, demonstrating confinement-mediated symmetry control. In panel (c), Bloch oscillations are monitored diabatically, with the system occupying superposed states in both the lower and upper bands. The transition from asymmetric to symmetric wave packets can be attributed to Landau-Zener (LZ) transitions occurring at unavoidable wave-packet touchings. These transitions arise from quantum superposition near avoided crossings, with the LZ transition probability determined by the interwave coupling strength \cite{Danga4}. The period of the topological drive protocol is comparable to the qubit dephasing timescale, and coherence is preserved within the modulation topological phase as the curvature of the confinement increases. Numerical analysis of non-adiabatic transitions at these wave-touchings and their impact on unconventional Bloch oscillations indicates the potential for tunable quantum transport applications. These approaches are also employed to investigate the relationship between irregular Bloch oscillations and tunneling parameters in a parabolic quantum wire subjected to biharmonic driving. This conclusion is corroborated by the consistency among experimental results, analytical theory, and simulations.

\subsection{Topological Transition and Chiral Interference}\label{subsec31}
\noindent

The parameters $E_k$ are designated as quasi-energies in this system, where $k = 1, 2$. These eigenvalues $E_k$ are mapped into the first Brillouin zone, satisfying $-\hbar\omega/2 < E_k < \hbar\omega/2$. In the quasi-energy basis $|\Phi_k(t)\rangle$, the transition probability $P_{|\downarrow\rangle \rightarrow |\uparrow\rangle}(t, t_0 )$ is expressed as follows.
\begin{equation}
P_{|\downarrow\rangle \rightarrow |\uparrow\rangle}(t, t_0 ) = \sum_{k, l}\exp(-i(E_k - E_l)(t - t_0 )/\hbar)A_k(t, t_0)A_l^{\ast}(t, t_0),
\end{equation}
where, $A_k(t, t_0 ) = \langle\uparrow|\Phi_k(t)\rangle\langle\Phi_k|\downarrow\rangle$. Eq. (22) provides the probability of a transition from the initial qubit state $|\downarrow\rangle$ to the final quantum state $|\uparrow\rangle$, incorporating quantum superposition summed over $l$. The result is subsequently summed over all final field states (summed over $k$). At this stage, contributions with different $k$ and $l$ values exhibit pronounced oscillations, thereby diminishing the overall transition probability. Modifying system parameters, such as the curved confinement $\Omega$ of the wire, can bring two quasi-energies close to degeneracy, $E_k = E_l$. Under these circumstances, the transition probability increases substantially. Therefore, the principal quantity of interest is the transition probability averaged over initial times $t_0$, while keeping the elapsed time $t - t_0$ fixed. This average can be directly obtained from Eq. (22), providing the required result
\begin{equation}
P_{|\downarrow\rangle \rightarrow |\uparrow\rangle}(t, t_0 ) = \sum_{k}\sum_{n, l}\Big|\langle\uparrow|\Phi^{(n-l)}k\rangle\Big|^2\Big|\langle\Phi^{(n-l)}k|\downarrow\rangle\Big|^2
\end{equation}
where calculations $\Phi_k^{(n)} = \frac{1}{T}\int^T_0\exp(in\omega{t})\Phi_k(t)dt$ are the Fourier components of the quasi-energy function. Eq. (20) is simply a generalization of the Rabi formula long used by workers in atomic and molecular beam resonance spectroscopy \cite{Kusch}. Considering only the resonant term in the $k$ and $n$ sum of Eq. (20) in Ref. \cite{Shirley}, the probability to go from a single initial Floquet state  $|\downarrow, 0\rangle$  to a final Floquet $|\uparrow,k\rangle$, summed over all $k$ in the RWA, is given by
\begin{equation}
P_{|\downarrow\rangle \rightarrow |\uparrow\rangle}(t, t_0 ) = \frac{|\Delta_{\text{eff}}^{n, m}|^2}{2E_{n, m}^2}\big[1 - \cos(E_{n, m})t\big]
\end{equation}

\begin{figure}
\begin{center}
	\includegraphics[width=14cm]{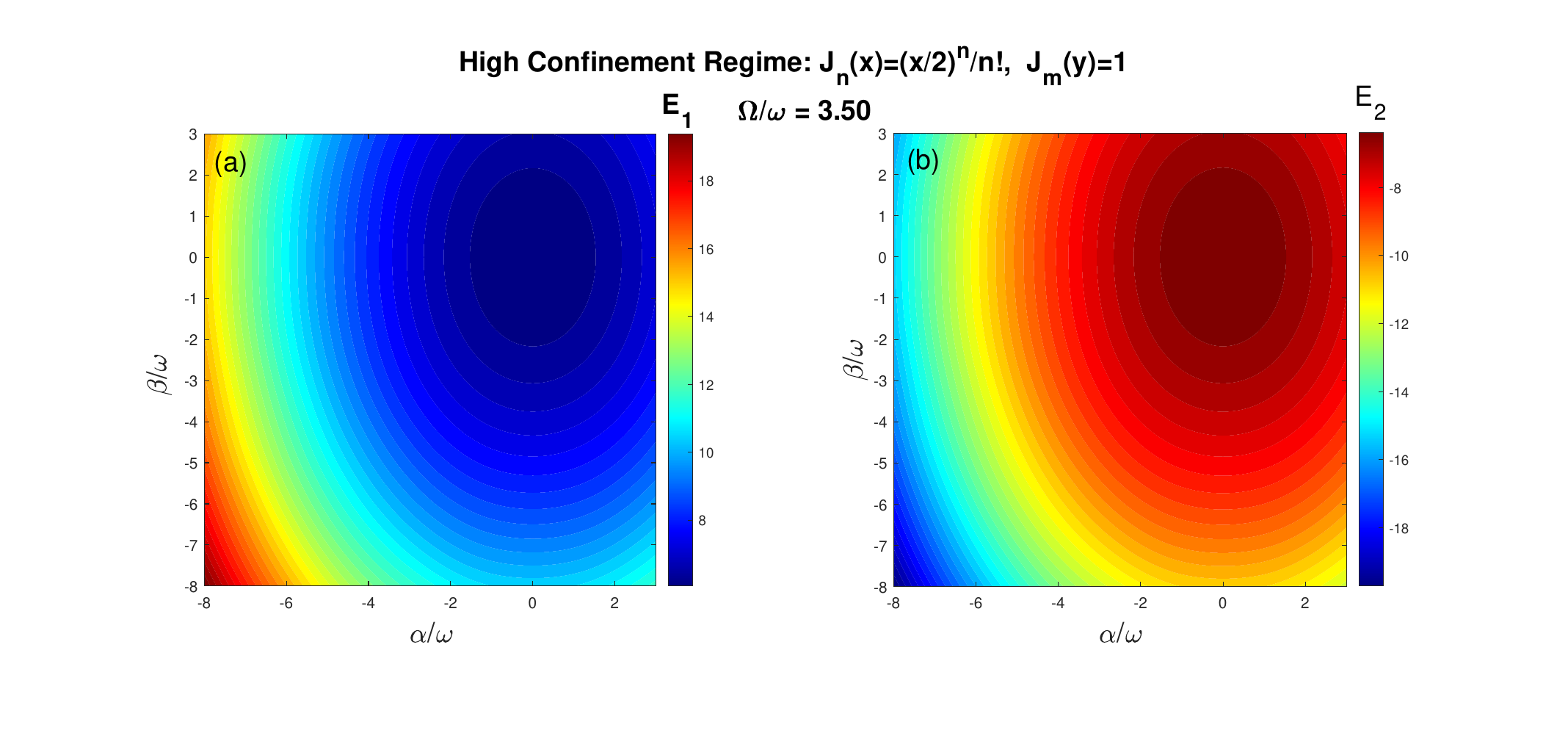}
	\includegraphics[width=14cm]{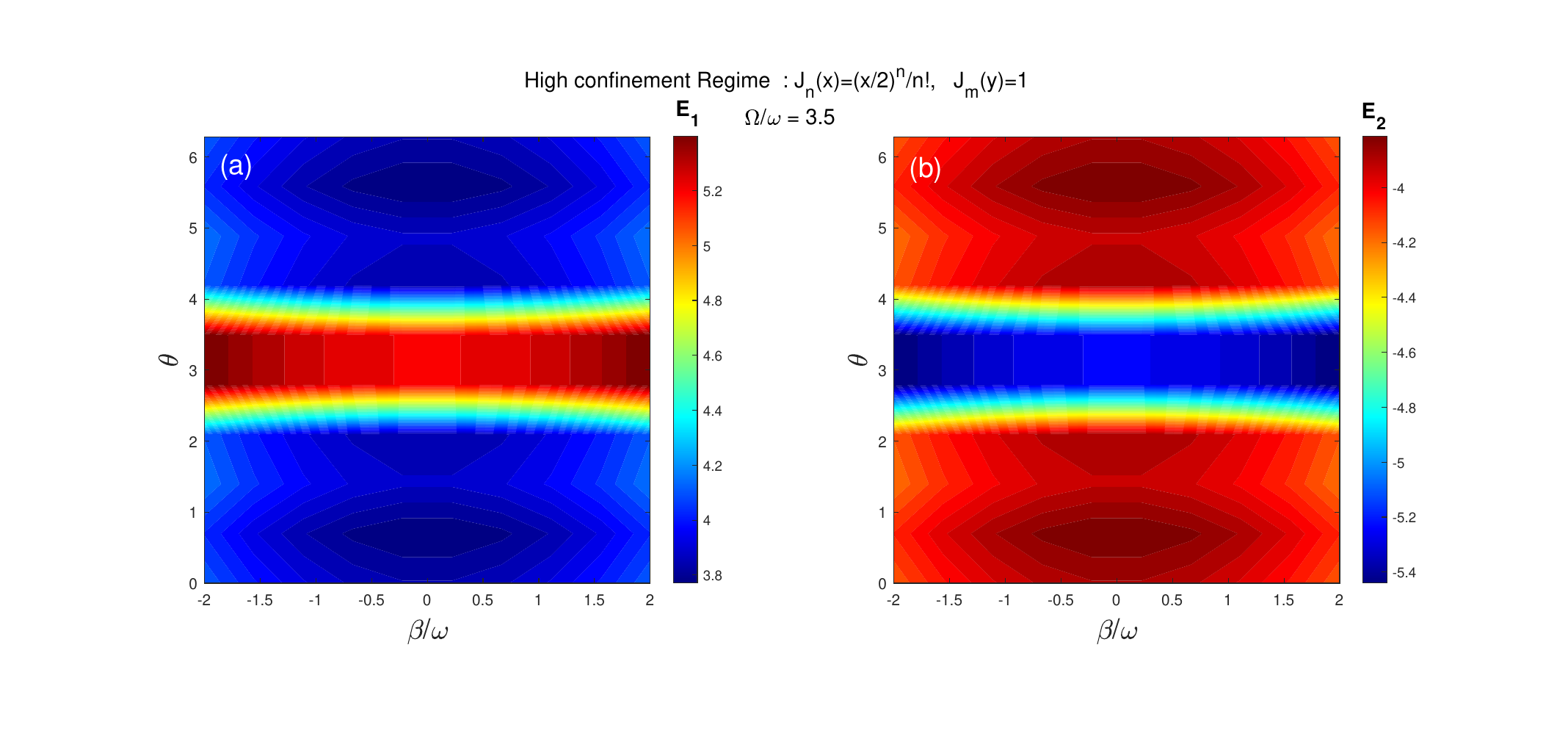}
\caption{(Color online) Quasi-energy level waveforms for $E_1$ and $E_2$ are presented as functions of $\alpha/\omega$, $\beta/\omega$, and in the $\beta$, $\theta$ plane under a high confinement regime ($\Omega/\omega = 3.5$). The drive symmetry parameter is set at $\theta = 0$ radians with mode parameters $n = 5$ and $m = 5$. Additional parameters are $\Delta/\omega = 0.47$ and $\alpha/\omega = 4.4$.}
\end{center}
\end{figure}

\begin{figure}
\begin{center}
	\includegraphics[width=14cm]{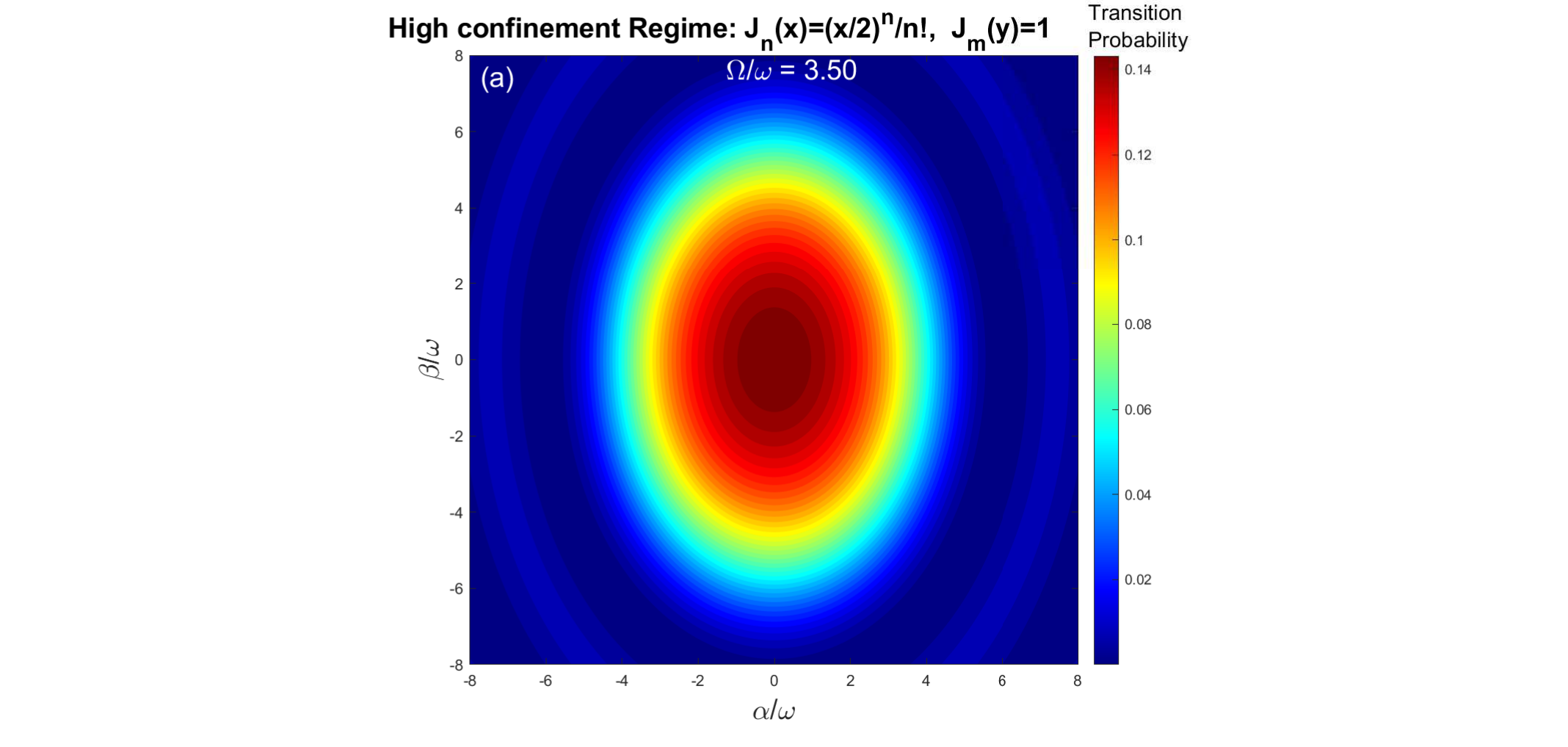}
	\includegraphics[width=14cm]{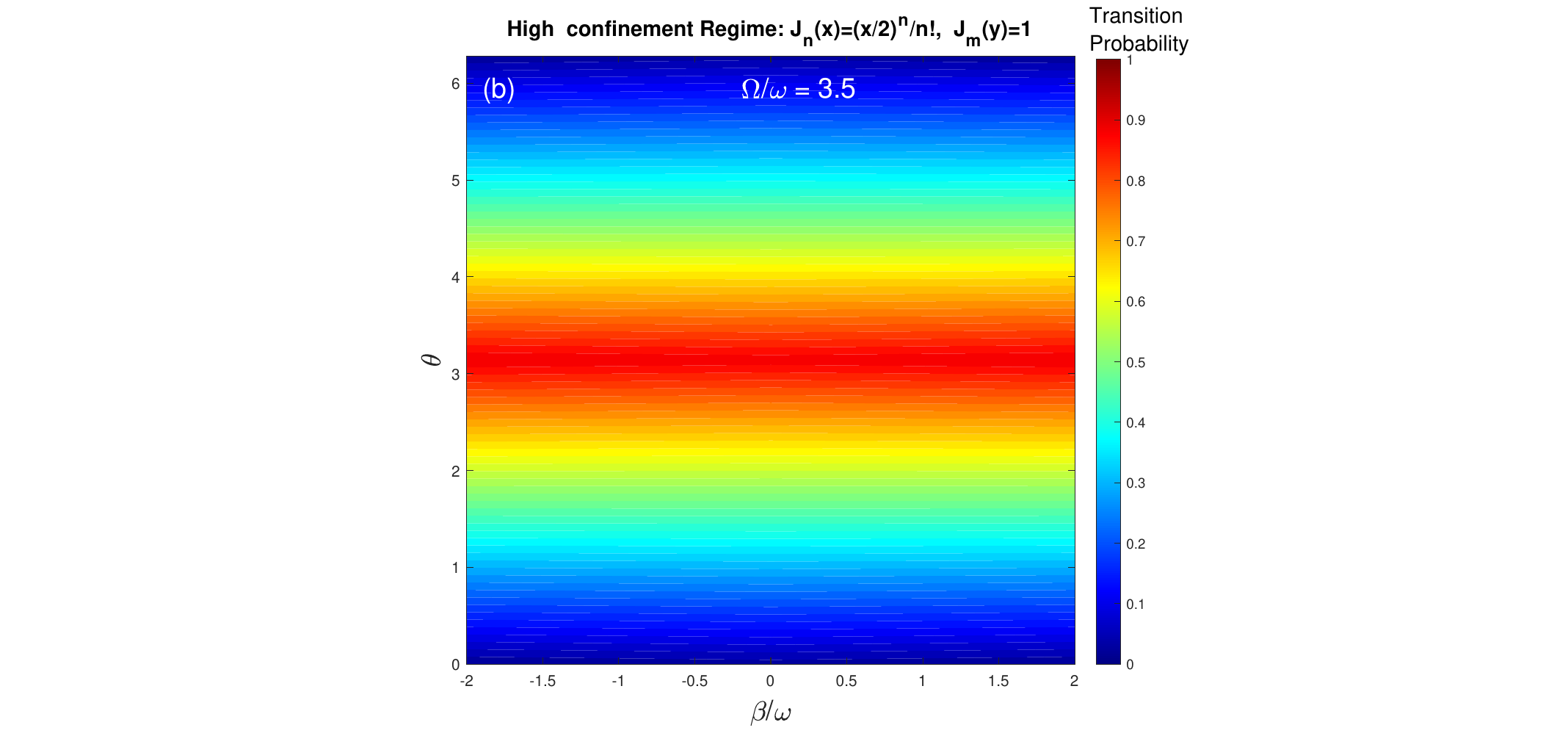}
\caption{(Color online) Transition probability waveform as a function of ($\alpha/\omega$, $\beta/\omega$), and the ($\beta/\omega$, $\theta$) plane in the high confinement regime ($\Omega/\omega = 3.5$), with the drive symmetry parameter fitted at $\theta = 0$ radians and mode parameters $n = 5$ and $m = 5$. The other parameter is $\Delta/\omega = 0.47$.}
\end{center}
\end{figure}

\subsection{Non-Abelian Geometric Phases and Holonomic Quantum  Computation}
\noindent

Non-Abelian geometric phases are typically represented as matrix-valued holonomies acquired by degenerate quantum states during cyclic evolution. These phases form the basis of holonomic quantum computation, enabling universal and intrinsically robust logical gate operations that are determined by the geometry of the evolution path rather than by precise interaction durations.

\subsubsection{High Confinement Regime (large $\Omega$)}
\noindent

High-confinement regimes, characterized by stringent spatial or potential boundaries, lead to increased energy-level splittings and modified wavefunction overlaps. These phenomena enable precise control over synthetic gauge fields and non-commutative geometric transformations, both of which are critical to the realization of fault-tolerant, robust holonomic quantum computation. In such regimes, spatial wavefunctions become more localized, resulting in heightened sensitivity to external potentials and increased overlap with neighboring states. This, in turn, alters the non-Abelian gauge potentials and influences the dynamics of geometric gates. Figures 2, 3, and 4 present the quantum behavior of a system under periodic driving or confinement, demonstrating the evolution of specific energy states and transition probabilities.

The validity of the RWA requires that the effective driving term remains significantly smaller than the driving frequency ($|\Delta_{\text{eff}}^{n, m}| \ll \omega$) and that the resonance is isolated, such that higher-order off-resonant terms average to zero over time. The effective coupling $\Delta_{\text{eff}}^{n, m}$ is determined by the Bessel functions of the first kind, ${J}_n\Big(\frac{\gamma_2}{\hbar\omega}\Big)$ and ${J}_m\Big(\frac{\gamma_3}{2\hbar\omega}\Big)$, which significantly influence the quasi-energy spectrum and transition probabilities.

A clear physical distinction must be made: a single two-level system (qubit) subjected to periodic driving generates an Abelian [U(1)] geometric phase, known as the Berry phase, rather than a non-Abelian phase. Non-Abelian geometric phases, which underpin Holonomic Topological Quantum Computation (HQC), require evolution within a degenerate subspace of dimension $N \geq 2$. Although the specific Hamiltonian considered cannot independently generate non-Abelian operations, the model remains a fundamental component of topological quantum approaches.

\subsubsection{Abelian Geometric Phase}
\noindent

When the parameters are varied adiabatically over a closed loop in parameter space, such that $\gamma_2$ and $\gamma_3$ are very small ($\frac{\gamma_2}{\hbar\omega} \ll 1$ and $\frac{\gamma_3}{\hbar\omega} \ll 1$), the Bessel functions can be approximated as $J_n(x) \approx (x/2)^n/n!$ and $J_0(y) \approx 1$. Under these conditions, the single quantum state acquires a standard $U(1)$ Berry phase. The state vector accumulates a complex phase factor $e^{i\gamma}\Big|_{i= 2, 3}$, while its orientation in Hilbert space remains unchanged. Unless a primary resonance ($m = 0$) is driven, the effective multiphoton Rabi frequency $\Delta_{\text{eff}}^{n, m}(\Omega, \theta)$ approaches zero \cite{Shevchenko, Ivakhnenko, Grifoni}. Consequently, the detuning term $\Delta$ dominates the negligibly small effective gap.

Colored waveguides in Fig. 2 illustrate the site-heterostructure magnetic quantum wire, which exhibits two quasi-energies of equal magnitude but opposite phase during uniform Floquet-Bloch oscillations. Adjusting system parameters in the high-confinement regime ($\Omega/\omega = 3.5$) or varying the phase $\theta$ can induce destructive interference in neighboring sites (Fig. 3). This adjustment lifts the band degeneracy, resulting in the formation of gaps between the colored flatband and dispersive bands at $\alpha/\omega = 0$. Determining the population probabilities of the system levels requires analysis of the quasi-energy spectrum. Previous studies \cite{Bukov, Forster, Danga, Danga2, Pachos, Pachos2} have shown that this approach accurately characterizes intermediate states of the driven system at optimal amplitude and enables identification of resonance transitions arising from the dynamics and crossing of quasi-energy levels. Generally, quasi-energy crossings significantly influence the population distribution among levels in complex quantum systems \cite{Izmalkov}. In this regime, the quasi-energies are confined within the minimum $\pm\Delta/2$ range (Fig. 2), and the maximum probability amplitude for reaching the excited state is nearly zero. The qubit remains predominantly in its initial state. The Floquet states of the Hamiltonian accumulate a standard U(1) Berry phase. The system dynamics become effectively frozen; the transition probability is negligible because the drive is insufficient to overcome the energy gap, as shown in Fig. 3.

Addressing the non-Abelian geometric phase requires grounding the discussion in fundamental quantum mechanics: a single, isolated two-level system is governed by the $SU(2)$ group and strictly possesses only an Abelian $U(1)$ geometric phase. Mathematically, it cannot independently support a non-Abelian phase.

\begin{figure}
\begin{center}
	\includegraphics[width=14cm]{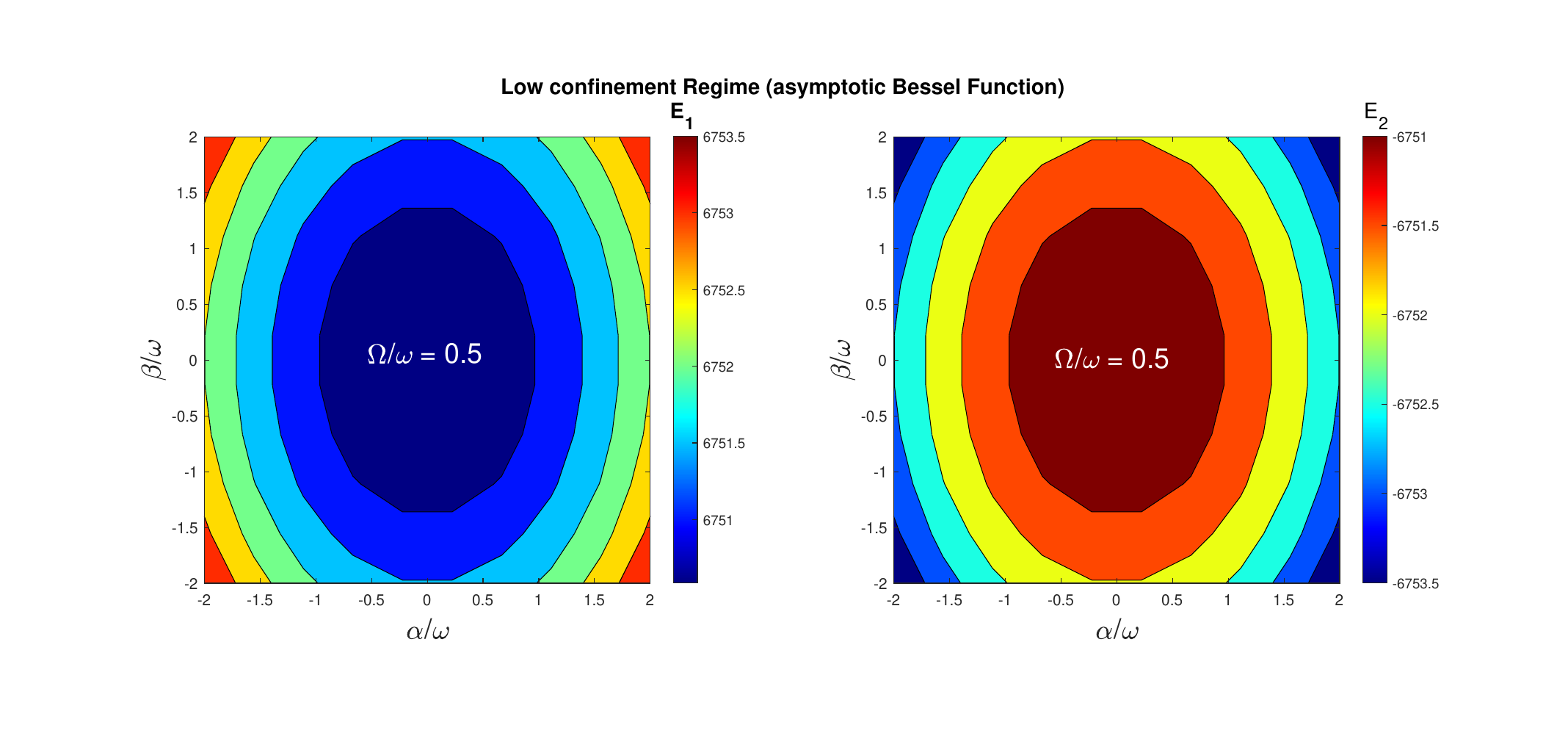}
\caption{(Color online) The quasi-energy level waveforms for $E_1$ and $E_2$ are shown in the low confinement regime, where $\Delta/\omega = 0.5$. The drive symmetry parameter is set at  $\theta = 0$ radians, with mode parameters $n = 5$ and $m = 5$. An additional parameter is $\Delta/\omega = 0.47$.}
\end{center}
\end{figure}

\begin{figure}
\begin{center}
	\includegraphics[width=14cm]{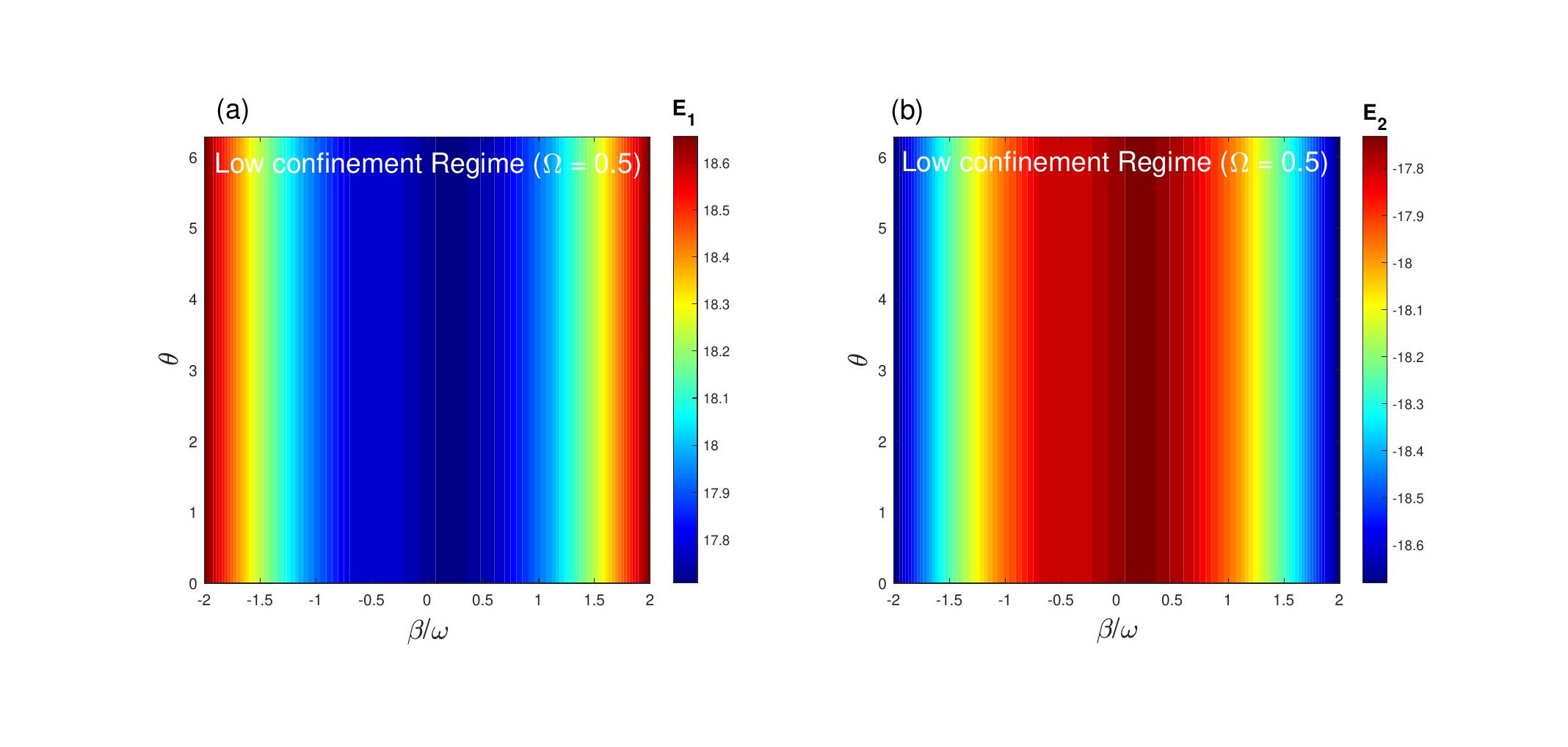}
\caption{(Color online) Quasi-energy level waveforms for $E_1$ and $E_2$ are presented in the low confinement regime, where $\Delta/\omega = 0.5$, on the contour plane defined by $\beta/\omega$ and $\theta$, with mode parameters $n = 5$ and $m = 5$. The other parameter is set as $\Delta/\omega = 0.47$.}
\end{center}
\end{figure}

\begin{figure}
\begin{center}
	\includegraphics[width=14cm]{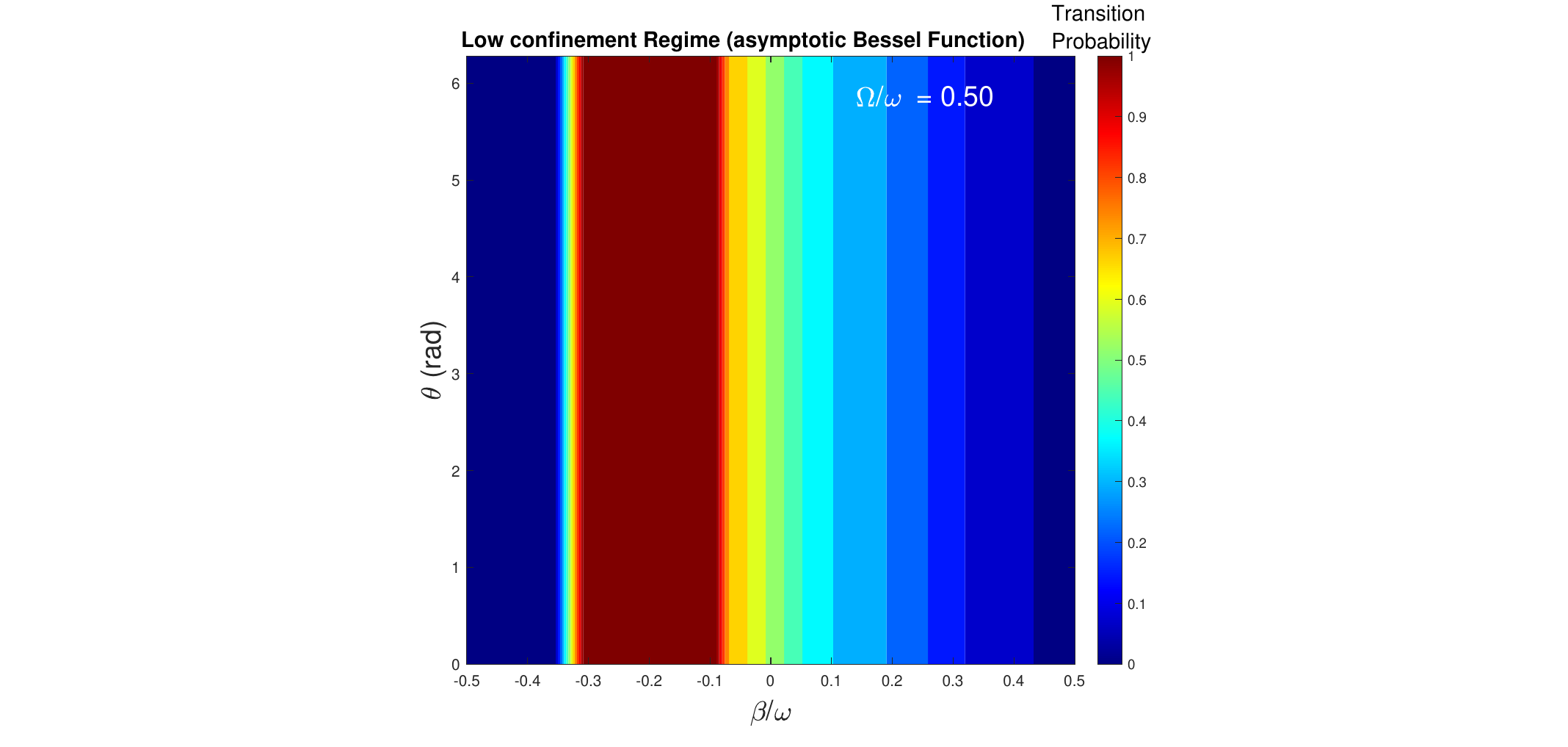}
\caption{(Color online) Transition probability waveforms for energy levels $E_1$ and $E_2$ are shown in the low confinement regime, where $\Delta/\omega = 0.5$, on a contour plane defined by $\beta/\omega$ and $\theta$, with mode parameters $n = 5$ and $m = 5$. Other parameters are set to $\Delta/\omega = 0.47$.}
\end{center}
\end{figure}

\begin{figure}
\begin{center}
	\includegraphics[width=14cm]{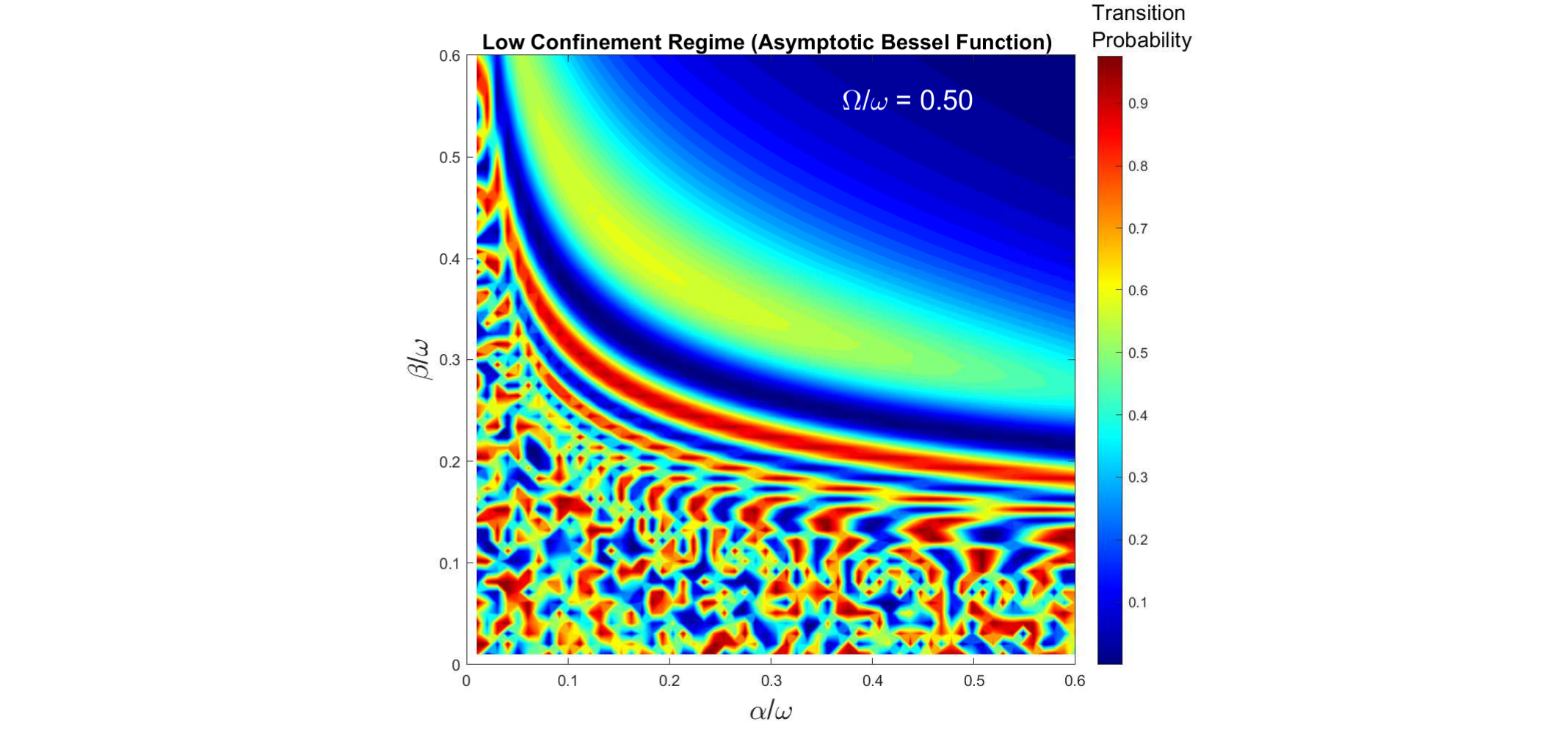}
\caption{(Color online) Transition probability waveforms for various levels $P_{|\downarrow\rangle \rightarrow |\uparrow\rangle}$ are shown in the low confinement regime ($\Delta/\omega = 0.5$), with the drive symmetry parameter set to $\theta = 0$ rad and mode parameters $n = 5$ and $m = 5$. Additional parameters include $\Delta/\omega = 0.47$.}
\end{center}
\end{figure}

\subsubsection{Low confinement Regime (large $\Omega$)}
\noindent

In low-confinement regimes, spatial wavefunctions spread out, increasing sensitivity to external potentials and neighbor-state overlap, which alters the non-Abelian gauge potentials and modifies geometric gate dynamics. Figures 5, 6, and 7 depict the quantum behavior of a system subjected to periodic driving or confinement, illustrating the evolution of specific energy states and transition probabilities. Figure 5 presents the quasi-energy level waveforms for the lower quantum states $E_{1}$ and $E_{2}$, demonstrating how periodic perturbations alter these energy values. Figure 6 provides the transition probability waveforms for the same $E_{1}$ and $E_{2}$ levels, indicating the likelihood of transitions between states over time. Figure 7 extends the transition probability analysis to additional energy levels in the low-confinement regime, where spatial restrictions are minimal, and wavefunctions exhibit greater spatial extension.

\subsubsection{Engineering a Degenerate Subspace (Path to Non-Abelian)}
\noindent

The realization of topological non-Abelian properties through Floquet engineering necessitates the coupling of two or more qubits or the inclusion of a third energy level in the Hamiltonian. Application of a high-frequency Floquet drive can dynamically suppress specific couplings, a process referred to as Coherent Destruction of Tunneling, where $J_0(x) = 0$. This approach facilitates the creation of a synthetic, multi-fold-degenerate dark subspace.

Non-Abelian phases require a degenerate, multi-dimensional subspace. Varying the parameters $\gamma_2$ and $\gamma_3$ in the Floquet Hamiltonian is the method by which this required degeneracy is engineered in larger systems.
 
In the low confinement limit, where $\Omega$ is very small, the system enters the ultra-strong driving regime. Here, the drive amplitude greatly exceeds the drive frequency ($\gamma_2 \gg \omega$, $\gamma_3 \gg \omega$), causing the Bessel functions to enter their asymptotic, highly oscillatory regime and decay as follows:
\begin{equation}
J_n(x) \approx \frac{2}{\pi{x}}\cos(x - \frac{n\pi}{2} - \frac{\pi}{4}).
\end{equation}                                                                            
The fundamental assumption of the RWA, that fast-oscillating terms can be neglected-completely breaks down in this regime. Off-resonant terms generate significant generalized Bloch-Siegert shifts, altering the resonance frequencies by amounts comparable to $\omega$. In this strongly driven regime, the analytical Floquet-RWA fails, and the system exhibits chaotic population transfer. As $\gamma_2$ and $\gamma_3$ increase, the Bessel functions oscillate. When the drive amplitudes are tuned such that the arguments coincide with the exact zeros of the Bessel functions (for example, $\gamma_2/\omega \approx 2.4$ where $J_0(x)=0$), the effective coupling $\Delta_{\text{eff}}^{n, m}(\Omega, \theta)$ vanishes, resulting in exact degeneracy of the two Floquet quasi-energies $E_1^{n, m}(\gamma_2, \gamma_3, \theta) = E_2^{n, m}(\gamma_2, \gamma_3, \theta)$ (see Fig. 5). Adiabatic modulation of these parameters in a closed loop near the degeneracy point confines the system to a two-dimensional subspace. Since $\bar{H}(t)$ is constructed from non-commuting Pauli matrices ($\sigma_x$, $\sigma_y$, $\sigma_z$), derivatives with respect to the parameters project onto the $SU(2)$ algebra. The structure of the second harmonic ($2\omega$) and the relative phase $\theta$ provide the necessary control parameters to manipulate the effective couplings in orthogonal directions (the $\sigma_x$ and $\sigma_y$ terms). Modulating these parameters traces a loop in the $SU(2)$ parameter space, thereby inducing non-Abelian geometric phases \cite{Wilczek, Zanardi, Nayak} on the quantum state. Figures 4 and 5 display the quasi-energy spectrum in the ($\alpha/\omega$, $\beta/\omega$) and ($\beta/\omega$, $\theta$) contour planes. Under low-confinement conditions ($\Omega/\omega = 0.5$), the spectrum exhibits symmetric interference fringes as the system transitions into the asymptotic Bessel-function regime.

The transition probability amplitude is highly sensitive to the precise values of $\gamma_2$ and $\gamma_3$. Since $\gamma_2$ and $\gamma_3$ appear in the denominator, the overall transition amplitude is suppressed as the driving strengths approach infinity (as shown in Fig. 5, due to the high-frequency drive averaging the transverse field to zero). When $\gamma_2$ and $\gamma_3$ are selected such that $\Theta_{n, m} = 0$, the quasi-energies collapse to exactly $\pm(\gamma_1 - m\omega)$, and the transition probability $P_{|\downarrow\rangle \rightarrow |\uparrow\rangle}$ becomes strictly zero, resulting in permanent localization of the particle, indicated by a blue band in Fig. 5 (Coherent Destruction of Tunneling). This observation is crucial for elucidating the emergence of Landau-Zener-St$\ddot{u}$ckelberg (LZS) interferometry transitions under the applied driving protocol. The control and measurement protocol involves observing coherent multiphoton resonances (Figs. 5, 6, 7) between discrete states to classify topological phases. Sudden configuration changes are observed, transitioning from an almost unaltered state (Figs. 5, 6, 7) to interband transitions while conserving the symmetric cell. As the curvature of the confinement increases, the beam intensity exhibits an asymmetric energy distribution. This behavior provides inherent robustness against fluctuations in drive amplitude, addressing a significant challenge in quantum control \cite{HuWu}. These results may be generalized to higher-dimensional systems along high-symmetry paths within the Brillouin zone, thereby offering a broader understanding of electronic transport in complex lattice structures \cite{Scully, Shirley}. Consequently, the number of scatterers increases, and the properties of the incoming and outgoing leads are strongly influenced by the geometry of the quantum wire, which in turn affects the transition probabilities in LZSM interferometry. The emergence and characteristics of irregular configuration-space bands in quantum-wire materials, which exhibit distinct Floquet Bloch-oscillation topologies, require further investigation.

\section{IMPLICATIONS  FOR QUANTUM TECHNOLOGIES}
\noindent

The proposed theoretical model establishes a foundational framework with substantial implications for quantum technologies. Its validity will be determined by comparison with experimental data. Beyond addressing fundamental phenomena, this platform offers direct avenues for practical applications. The scope of discussion is extended to encompass specific proposals for experimental implementation, scalability, robustness evaluation, and prospective research directions.

\subsection{Proposal for application in GaAs/AlGaAs Semiconductor nanowire Heterostructures}

\begin{figure}
\begin{center}
	\includegraphics[width=14cm]{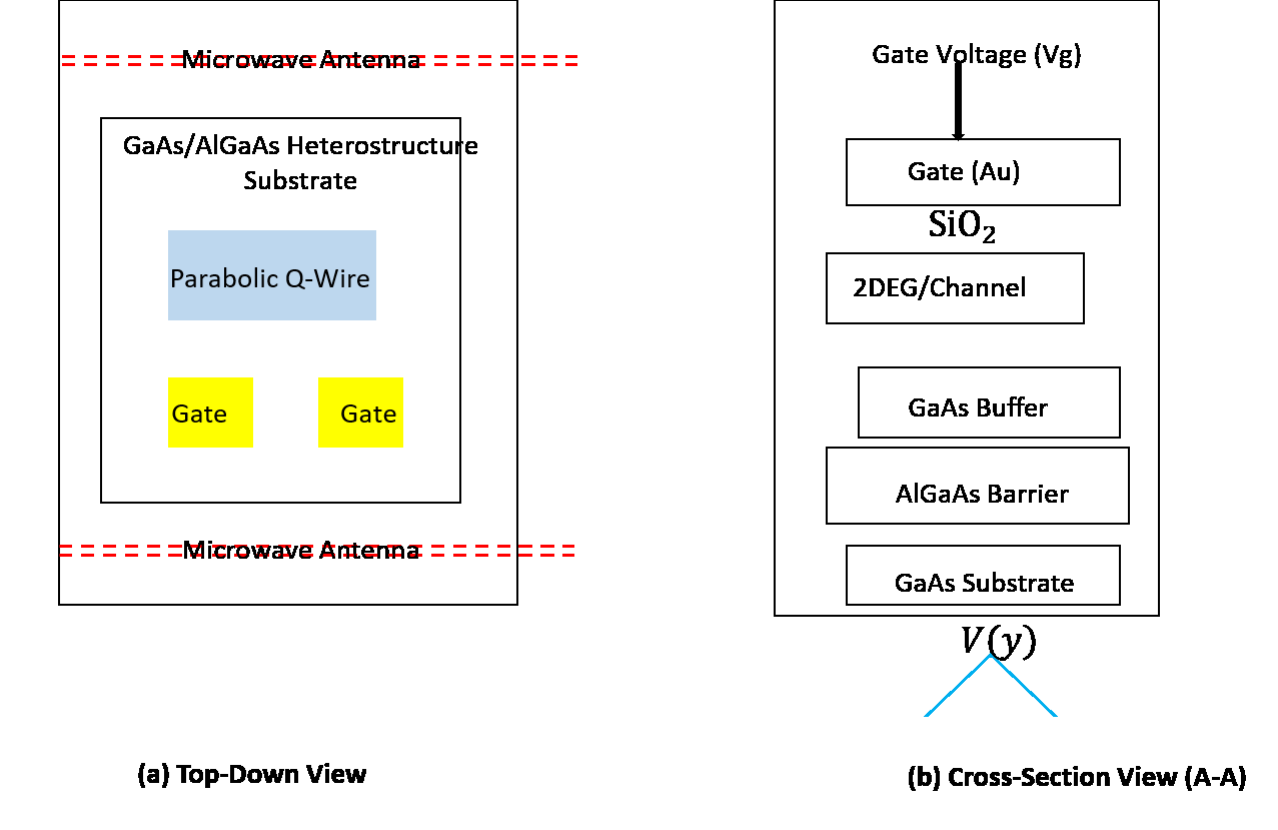}
\caption{Proposed Experimental Device Schematic: a) Top-Down View b) Cross-Sectional View (A-A). }
\end{center}
\end{figure}

The predicted effects are experimentally accessible using current nanofabrication and microwave control technologies. Surface gates define a one-dimensional parabolic quantum wire with a confinement energy $\hbar\Omega$ that is tunable from approximately 1 to 10 meV via gate voltages. Integrated on-chip microwave antennas deliver a biharmonic drive at frequencies $\frac{\omega}{2\pi}$ near 10 GHz, with amplitudes $\alpha$ and $\phi$ corresponding to magnetic fields of 1-10 mT and a phase $\theta$ controlled electronically. In high-mobility $\text{GaAs/AlGaAs}$ core-shell nanowires \cite{Lauhon, Funk, Morkotter, Fickenscher} (see Fig. 8(a)), a quasi-one-dimensional electron gas forms at the radial heterointerface. The spatial electrostatic potential $V(y)$ (see Fig. 8(b)) determines the parabolic confinement strength $\Omega$.

Spin state readout is achievable via Pauli spin blockade in an adjacent quantum dot or via integrated superconducting resonators for dispersive readout. This setup provides full control over the relevant parameters ($\Omega$, $\omega$, $\alpha$, $\beta$, $\phi$), enabling the mapping of phase diagrams and the observation of confinement-induced topological transitions, Floquet-Bloch oscillations, and geometric phases. This approach constitutes an implementation of Floquet engineering within a semiconductor platform.
  
When on-chip microwave antennas deliver static magnetic fields and biharmonic electromagnetic pulses, with $\alpha$ representing the continuous drive amplitude and $\beta$ the alternating pulse amplitude, the parameters scale with confinement as follows: static detuning bias $\gamma_1 = 1+ \frac{\alpha^2}{\Omega^2} + \frac{\beta^2}{4\Omega^2}$, primary harmonic amplitude $\gamma_2 = \frac{\alpha\cdot\beta}{\Omega}$, and secondary harmonic amplitude $\gamma_3 = \frac{\beta^2}{\Omega}$. Sweeping the contour plane ($\alpha$, $\beta$) under a fixed physical gap $\Delta = 40$ meV and microwave frequency $\frac{\omega}{2\pi} = 10$ GHz reveals distinctly different structural behaviors in the energy landscape, depending on the ratio of the drive fields to the confinement $\Omega$.
\begin{figure}
\begin{center}
	\includegraphics[width=14cm]{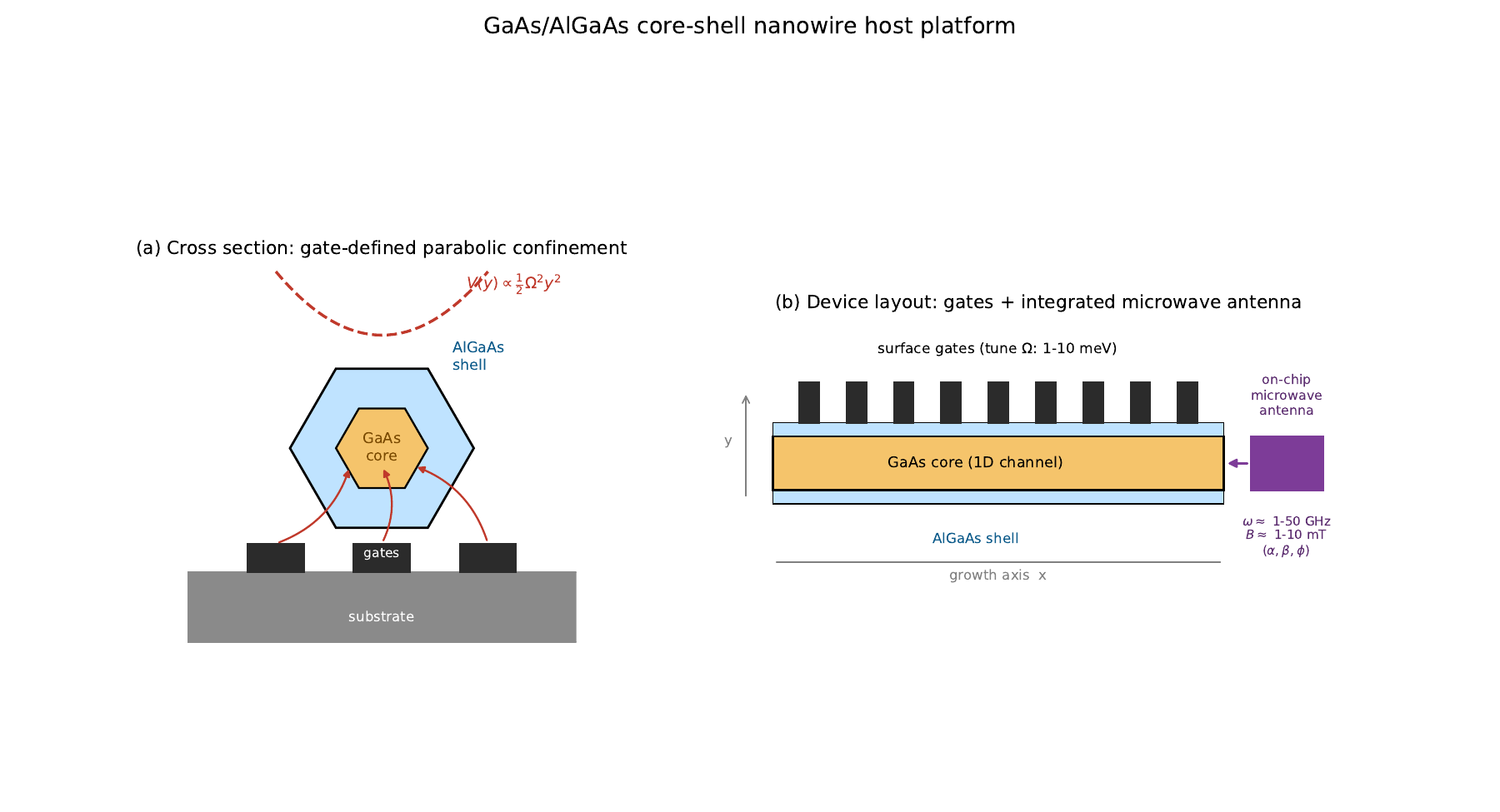}
\caption{(Color online) The figure illustrates a GaAs/AlGaAs core-shell nanowire host platform, a semiconductor nanostructure comprising a gallium arsenide (GaAs) core surrounded by a wider-bandgap aluminum gallium arsenide (AlGaAs) shell. This architecture enhances electronic passivation, enables quantum confinement, and offers a stable environment for nanoscale optoelectronic and quantum devices. Panel (a) presents a cross-sectional view of gate-defined parabolic confinement, while panel (b) depicts the device layout, including gates and an integrated microwave antenna.}
\end{center}
\end{figure}

\begin{figure}
\begin{center}
	\includegraphics[width=14cm]{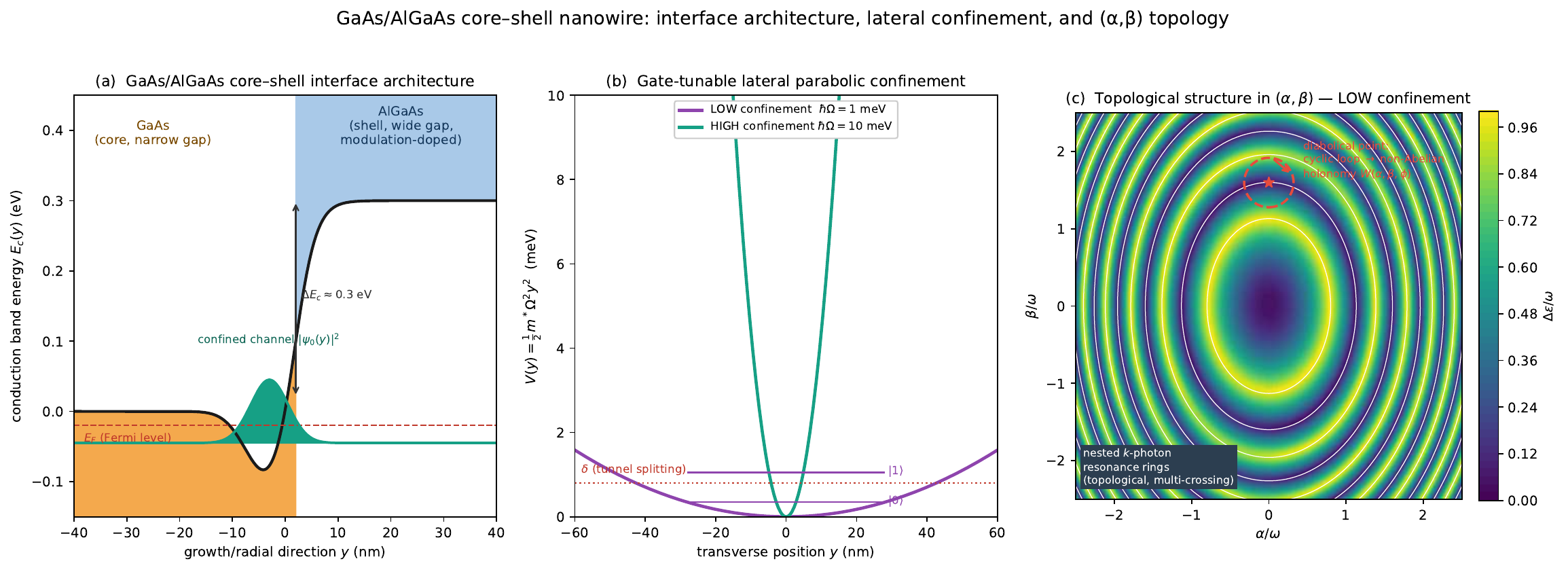}
\caption{(Color online) Panel (a) illustrates the GaAs/AlGaAs  core-shell nanowire with surface split-gates, which define the tunable parabolic potential $V(y)$. Panel (b) depicts two on-chip microwave antennas that deliver inhomogeneous ($\alpha\sin\omega{t}$) and homogeneous ($\beta\cos\omega{t} + \phi$) drive tones. The GaAs/AlGaAs core-shell nanowire interface architecture is shown, with the conduction band energy $E_c(y)$ plotted as a function of the growth (radial) direction y. The gate-tunable confinement curve is presented for $\hbar\Omega$ = 1 meV and $\hbar\Omega$ = 10 meV, with the $\delta$-scale tunnel splitting indicated. Panel (c) displays the topological features directly associated with Floquet plots, including nested $\gamma_1 = k\omega/2$ resonance rings, a marked diabolical (gap-closing) point, and a cyclic loop in ($\alpha$, $\beta$, $\phi$) whose transport generates the non-Abelian properties.}
\end{center}
\end{figure}

 \begin{figure}
\begin{center}
	\includegraphics[width=14cm]{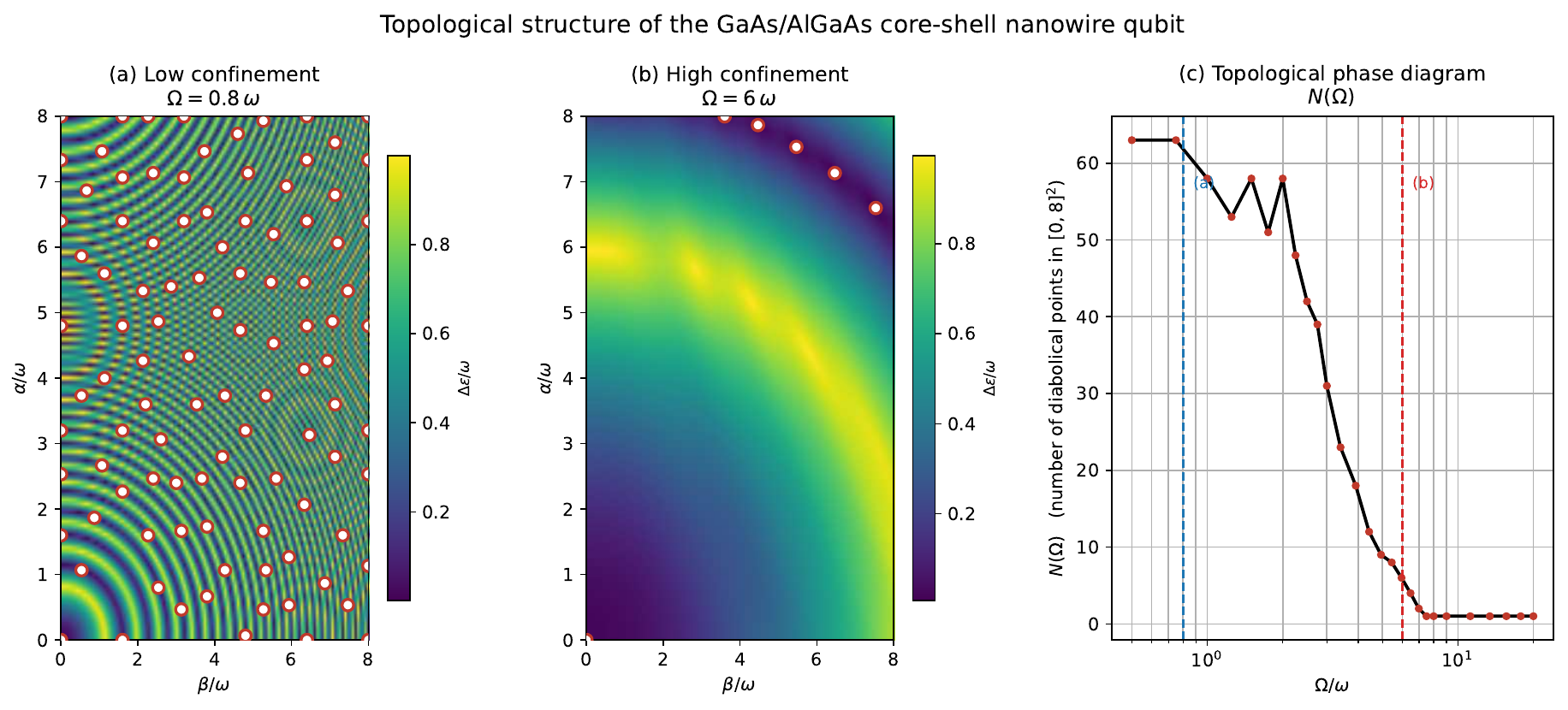}
\caption{(Color online) Panel (a): Sixty-three white dots indicate candidate diabolical points in the driven qubit's Floquet spectrum, corresponding to locations in the microwave-drive-amplitude plane where the multi-photon Rabi coupling nearly vanishes at a $k$-photon resonance. Panel (b): Under the same window and threshold, but with high confinement, only five diabolical points are observed. Panel (c): The quantitative relationship between confinement and topology is illustrated. $N(\Omega)$ represents the number of resonance rings, and thus candidate diabolical points, within a fixed, realistic drive-amplitude window.}
\end{center}
\end{figure}

In Figure 11, panel (a) displays 63 white dots that indicate candidate diabolical points in the driven qubit's Floquet spectrum. These points correspond to positions in the microwave-drive-amplitude plane where the multiphoton Rabi coupling approaches zero at a k-photon resonance. These points are determined by thresholding the exact numerical gap map, not by manual placement. The near-regular diagonal banding beneath the dots, visible as faint periodic ripples running from lower-left to upper-right, arises from interference fringes produced by many overlapping LZS resonances. This pattern is the deep-strong-driving analog of the sparser fringes observed at $\Omega = 5\omega$, but with more resonance orders superimposed. This regime is too complex for practical holonomic gate design, as the abundance of nearby diabolical points means that any adiabatic loop of reasonable size is likely to intersect several of them, complicating the isolation of a specific, well-characterized non-Abelian gate and increasing the risk of generating an uncontrolled mixture. Panel (b): Using the same window and threshold but with high confinement, only five diabolical points are present. The gap floor is ubiquitous, leaving nothing to encircle, which defines a trivial operating regime in the topological sense. Here, the ridge is shifted to the edge of the accessible ($\alpha$, $\beta$) window and reduced to a few crossings. Panel (c): This panel quantitatively illustrates the principle that confinement controls topology. $N(\Omega)$ denotes the number of resonance rings, and thus candidate diabolical points, within a fixed, realistic drive-amplitude window. Interpreted as a device design chart, the steep decline in $N(\Omega)$ between $\Omega/\omega \approx 1$ and $\Omega/\omega \approx 7$ marks the transition from an uncontrollable number of resonances to a manageable few. This transition is a crossover rather than a strict phase transition, as $N(\Omega)$ decreases smoothly (apart from grid sampling noise), with each diabolical point created or destroyed individually as $\Omega$ varies. The plateau at $N(\Omega) = 1$ for $\Omega/\omega \gtrsim 8$ should be noted: $N(\Omega)$ does not immediately drop to zero, as at least one residual near-degeneracy persists near the edge of any finite ($\alpha$, $\beta$) window, provided $\alpha$ and $\beta$ can become sufficiently large relative to $\Omega$, since $\gamma_1$, $\gamma_2$, and $\gamma_3$ never vanish exactly. When interpreted as a phase diagram, $\Omega$ (the gate voltage on the GaAs/AlGaAs core-shell nanowire) serves as a continuous parameter that determines the number of protected degeneracies, or independent holonomic-gate ``stations" accessible. This provides the concrete, testable basis for the term ``confinement-induced topological transition" used throughout this work.

\subsection{Non-Abelian holonomy and the topological LZ transition}
\noindent

\begin{figure}
\begin{center}
	\includegraphics[width=14cm]{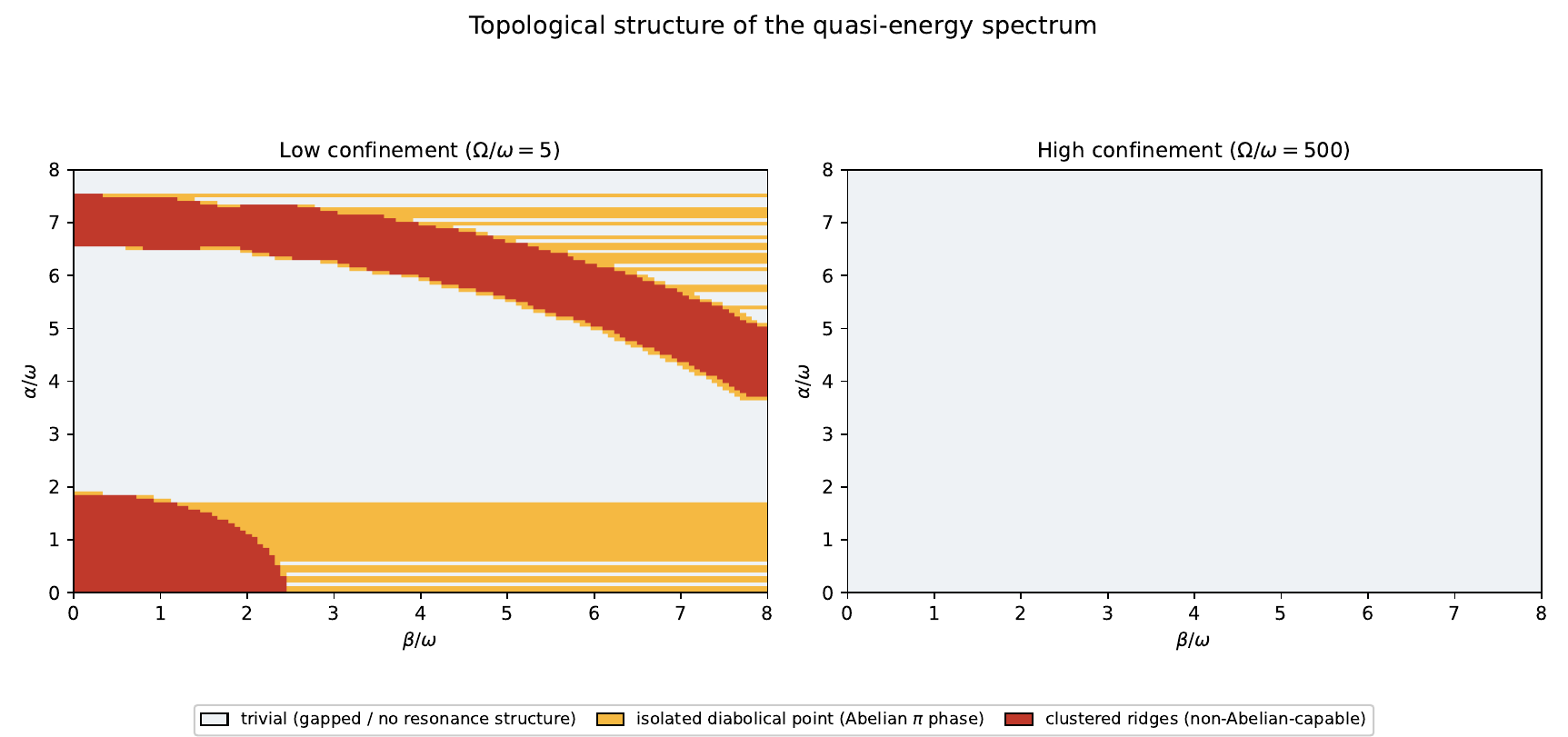}
    \includegraphics[width=14cm]{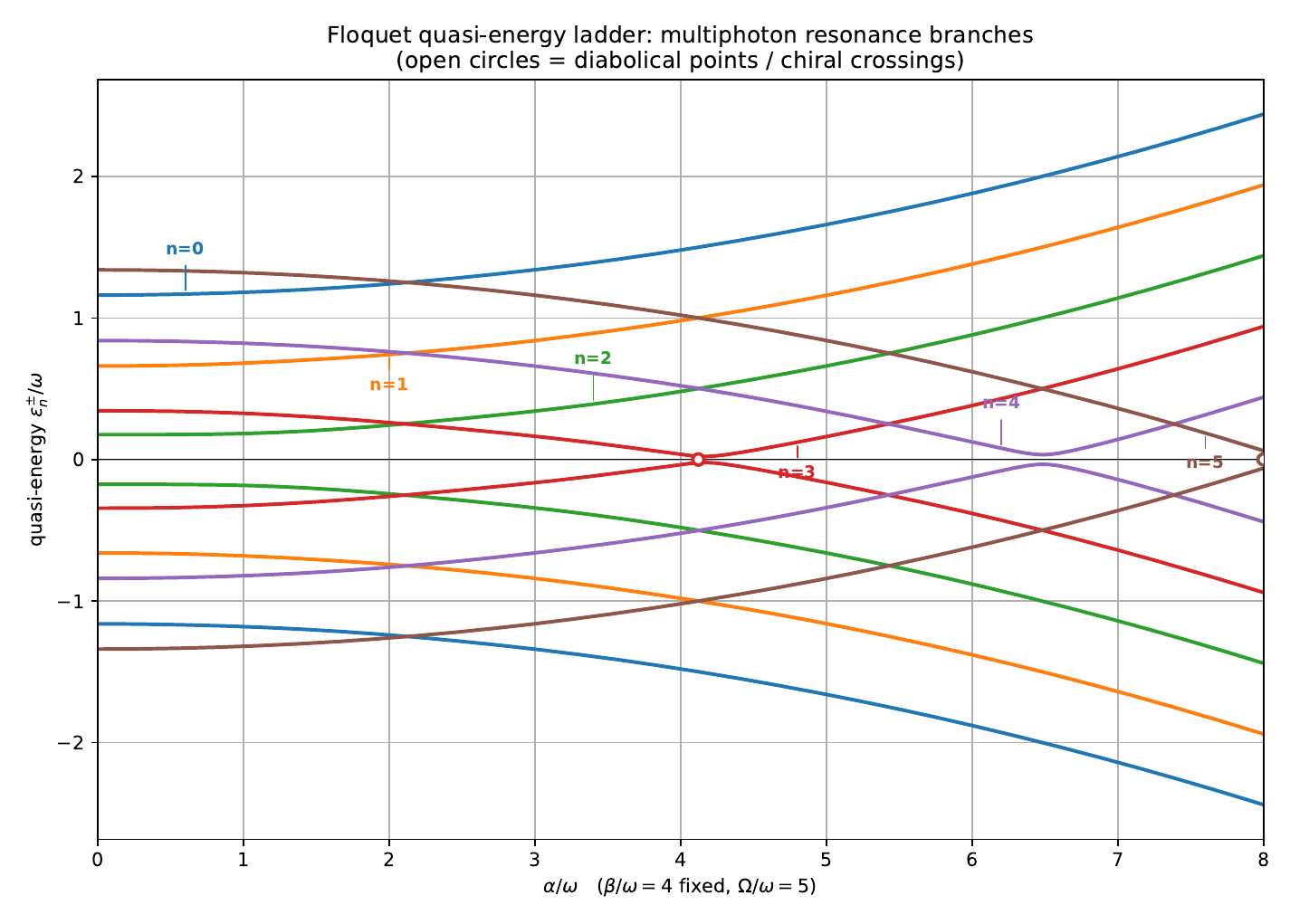}
\caption{(Color online) The first two graphs present the topological structure of the quasi-energy in both the low-confinement regime ($\Omega = 5\omega$) and the high-confinement regime ($\Omega = 500\omega$). The third graph displays a Floquet quasi-energy ladder, which characterizes the periodic, photon-spaced energy spectrum of a quantum system under a time-periodic field. Multiphoton resonance branches arise at the intersections of these energy ladders, leading to level hybridization and sideband-selective transitions as the field intensity and driving frequency change.}
\end{center}
\end{figure}

Fig. 12 illustrates topological structure of the quasi-energy: In the low-confinement regime ($\Omega = 5\omega$), the grey region (trivial) indicates that h-sweeps occur, but the resulting gap remains open; this corresponds to ordinary adiabatic following, with $P_{LZ} \approx 0$ and negligible accumulated phase. The orange region (isolated diabolical point) features a single near-zero-gap crossing, where exactly one LZ passage becomes non-adiabatic. Looping in the ($\alpha/\omega$, $\beta/\omega$) parameter space around one of these points results in the quantized Abelian $\pi$ Berry phase, which is the mechanism underlying the individual St$\ddot{u}$ckelberg fringes observed in earlier interference maps. The red region (clustered ridges) contains several near-zero-gap valleys that are in close proximity or cross, leading to multiple non-adiabatic LZ passages within a small loop. In multiphoton resonance branches, the open circle denotes a diabolical point or chiral crossing. This region allows a loop to enclose more than one diabolical point simultaneously, which is the geometric condition required for the transport to exhibit a genuinely matrix-valued (non-Abelian) holonomy, rather than a sum of independent scalar phases. The red region is concentrated at small $\alpha$, $\beta$ and along the bent ridge near $\alpha/\omega \approx 5-7$, indicating that large drive amplitudes are not necessary; instead, the crowding of resonance order ($n$) near $\gamma_1 \sim$ a few times $\omega/2$ is the critical factor. In the high-confinement regime ($\Omega = 500\omega$), increasing $\Omega$ suppresses $\gamma_2$ and $\gamma_3$ (with $\gamma_2, \gamma_3 \propto 1/\Omega$), which collapses the sweep amplitude. Without significant modulation of $h(t)$, repeated passages and avoided crossings are absent; the computed ``gap"  is simply the static tunnel splitting, which remains constant across the entire plane. Physically, this panel corresponds to the tight red parabola in the device schematic: increasing confinement eliminates access to LZS interference, diabolical points, and both Abelian and non-Abelian holonomy.

The exact numerical results are presented in this section. Since $H(t)$ is traceless, the one-period propagator $U(t)$, with $T=2\pi/\omega$, belongs to $SU(2)$; its eigenvalues are $e^{\pm{i\phi}}$ with
\begin{equation}
\phi = \arccos[\frac{1}{2}R_eT_rU(T)],  \phantom{...}  \Delta\varepsilon = \varepsilon_{+} - \varepsilon_{-} = \frac{2\phi}{T} 
\end{equation}
This operator is diagonalized numerically below using exact, stepwise propagation without approximations. The Bessel or multiphoton form can also be obtained analytically. The Jacobi-Anger expansion of $e^{-i\phi(t)}$ in equation (19) yields a Fourier series
\begin{equation}
e^{-{i}\phi(t)} = \exp{-{i}K(\gamma_2, \gamma_3)}\cdot\sum_{p, q = -\infty}^{+\infty}\Theta_{p, q}{{J}_p\Big(\frac{\gamma_2}{\hbar\omega}\Big)}\cdot{{J}_q\Big(\frac{\gamma_3}{2\hbar\omega}\Big)}e^{-{i}[\gamma_1 + (p + 2q)\omega]t}.
\end{equation}
From this equation, the exact expression for the phase factor associated with the $p^{th}$ and $q^{th}$ harmonics of the driving field is $\Theta_{p, q} = i^p\cdot{e^{i{q}\phi}}$. By collecting terms with $p + 2q = -k$ (integer photon number) and truncating to the resonant term $2\gamma_1 \cong k\omega$ (standard rotating wave approximation in the dressed frame), the multiphoton LZSM gap is obtained.
\begin{equation}
\Delta\varepsilon(\gamma_1, \gamma_2, \phi) = \pm\frac{1}{2}\sqrt{(\gamma_1 - k\omega)^2 + |\delta^{p, q}_{\text{eff}}|^2}.
\end{equation}
where, $\delta^{p, q}_{\text{eff}} = \delta\cdot\sum_{p, q = -\infty}^{+\infty}\Theta_{p, q}{{J}_p\Big(\frac{\gamma_2}{\hbar\omega}\Big)}\cdot{{J}_q\Big(\frac{\gamma_3}{2\hbar\omega}\Big)}$, where $\delta = \gamma_1 - k\omega$ represents the physical gap as defined in Eq.(18).

\subsubsection{Transition probability}
\noindent

Within a single rotating wave approximation branch, the stroboscopic (period-to-period) transition probability follows the standard Rabi-type Landau-Zener-St$\ddot{u}$ckelberg-Majorana (LZSM) formula.
\begin{equation}
P_k(t) = \frac{|\delta^{p, q}_{\text{eff}}|^2}{\Delta\varepsilon^2_k}\sin\big(\frac{\Delta\varepsilon_k}{2}t\big)
\end{equation}
The exact one-period value, $P = |U_{10}|^2$, without RWA truncation, is evaluated below.

At each multiphoton resonance, defined by $\gamma_1 = k\omega/2$, the two dressed quasi-energy branches become nearly degenerate. These are referred to as the ``multi-crossing" points. Adiabatic encirclement of such a point in ($\alpha$, $\beta$, $\phi$) space results in the accumulation of a geometric (Berry) phase, which is quantized by the winding of $\phi$(t). When a family of nearly degenerate Floquet sidebands is retained near a crossing, rather than a strictly isolated two-level subspace, cyclic transport generates a Wilczek-Zee connection. The holonomy matrices associated with this connection may not commute for loops encircling different crossings, providing the non-Abelian structure necessary for holonomic gates. Since $\gamma_2 = \alpha\cdot\beta/\Omega$ is odd under $\alpha \rightarrow -\alpha$ or $\beta \rightarrow -\beta$, while $\gamma_1$ and $\gamma_3$ are even, the resulting interference pattern exhibits exact point (inversion) symmetry, ($\alpha$, $\beta$) $\rightarrow$ (-$\alpha$, -$\beta$), but lacks mirror symmetry. This property defines the chirality known as ``chiral interference".

\subsubsection{Numerically exact Floquet check (contour maps)}
\noindent  

We diagonalized the exact one-period monodromy matrix (500-slice stepwise propagator, no RWA) on ($\alpha/\omega$, $\beta/\omega$) plane $\in$ [-2.5, 2.5], for $\delta = 0.4\omega$, $\phi =\pi/3$, at $\Omega/2\pi$ = 5 GHz (low confinement) and $\Omega/2\pi$ = 500 GHz (high confinement). White lines mark the  $k$-photon resonance rings $\gamma_1 = k\omega/2$.

\begin{figure}
\begin{center}
	\includegraphics[width=14cm]{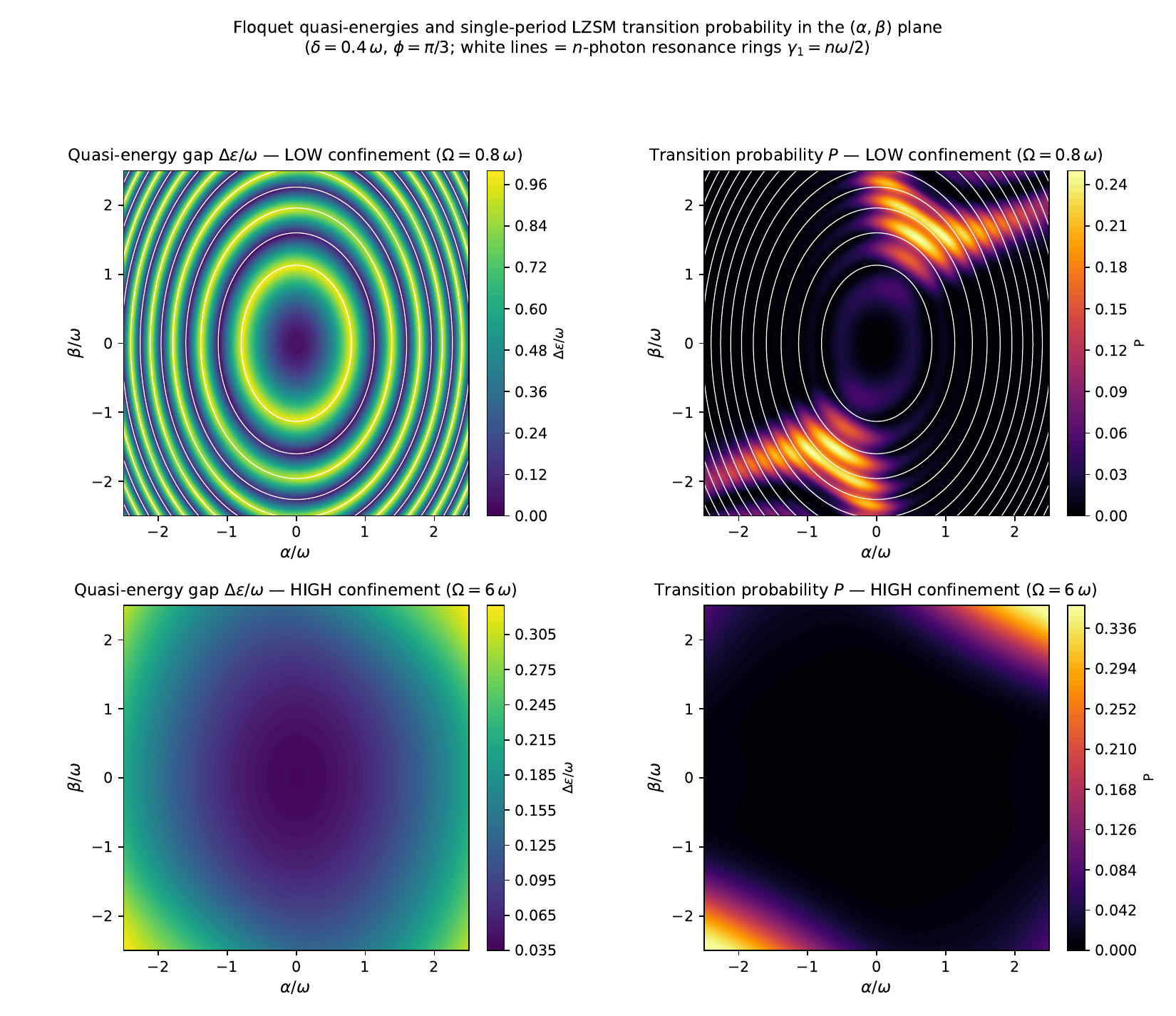}
\caption{(Color online) Quasi-energy and transition probability spectra at low ($\Omega/2\pi$ = 8 GHz) and high confinement ($\Omega/2\pi$ = 60 GHz) regimes, with fixed parameters $\delta/2\pi$ = 4 GHz, $\omega/2\pi$ = 10 GHz, and $\phi$ = $\pi/3$.}
\end{center}
\end{figure}
These results provide quantitative confirmation of the underlying physics. In the low-confinement regime ($\Omega/2\pi$ = 5 GHz), as shown in Fig. 12, the system develops near-gap-closing (diabolical) points that are nested as concentric resonance rings. In contrast, the high-confinement regime remains strictly gapped over the same drive-amplitude window, indicating a confinement-induced topological LZ transition. Inversion symmetry is exact in both regimes, as it is an exact symmetry of $H_{\text{eff}}$. However, only in the presence of diabolical points does encircling them acquire topological significance \cite{NadjPerge, RLi, Oreg, Lutchyn}. The combination of this symmetry with energetically significant mirror-symmetry breaking leads to a transition from a smooth, quasi-symmetric LZSM fringe pattern (high confinement) to a chiral, ring-nested interference pattern (low confinement) observed in the transition-probability panels.

\begin{figure}
\begin{center}
	\includegraphics[width=14cm]{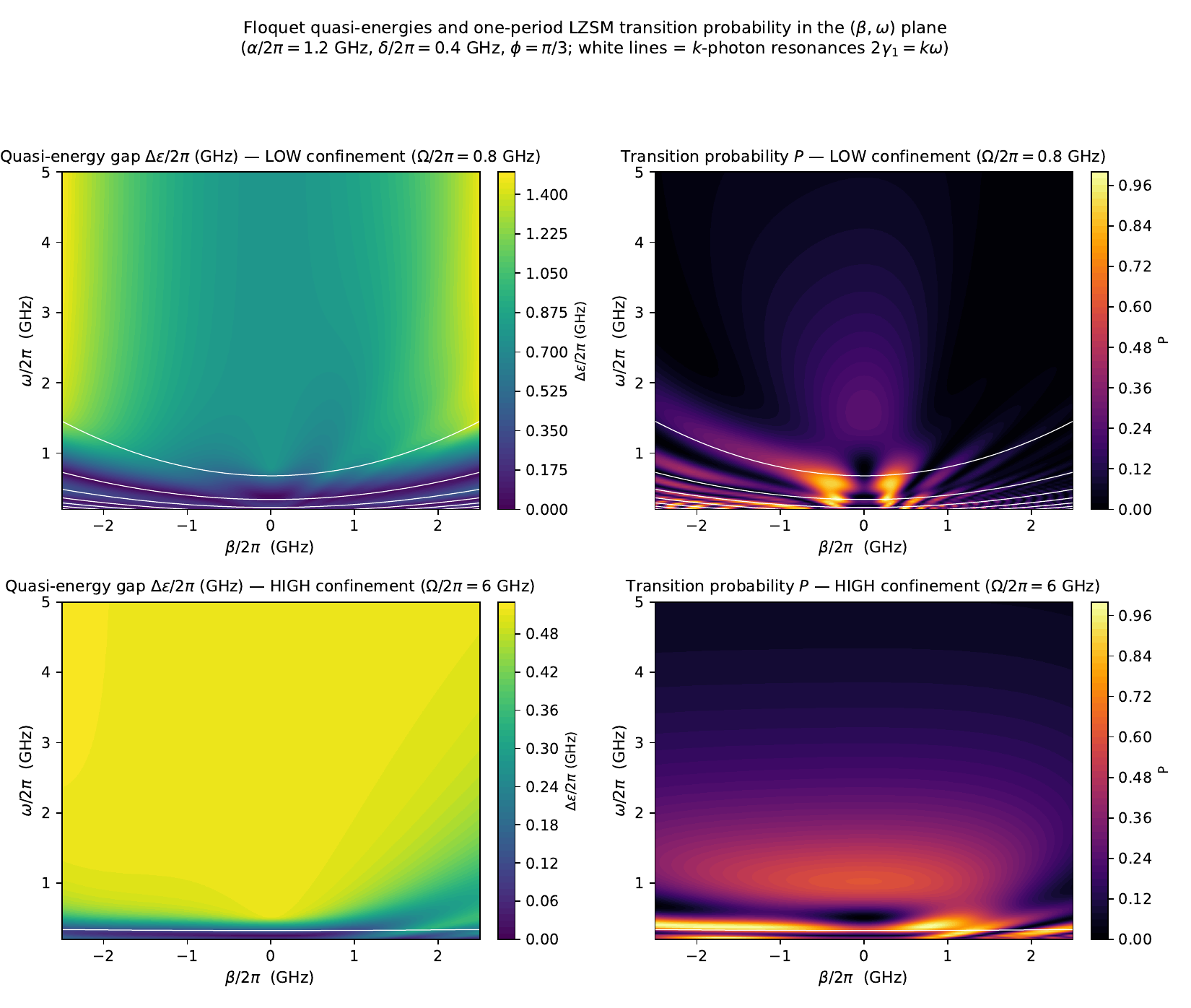}
\caption{(Color online) Floquet quasi-energies and one-period LZSM transition probabilities in the ($\beta/2\pi$, $\omega/2\pi$) plane are presented for both low ($\Omega/2\pi$ = 0.8 GHz) and high confinement regimes ($\Omega/2\pi$ = 6 GHz), with fixed parameters $\delta/2\pi$ = 0.4 GHz, $\alpha/2\pi$ = 1.2 GHz, and $\phi$ = $\pi/3$.}
\end{center}
\end{figure}

The static parameter $\gamma_1$ remains fixed near unity. In the quasi-energy contour plane, the spectrum displays highly linear, parallel, and stable energy bands. Avoided level crossings are extremely narrow due to significant suppression of effective cross-coupling.

In the high-confinement regime (see Fig. 13, with $\Omega/2\pi$ = 6 GHz), the quasi-energy surface remains strictly gapped due to strong suppression of effective cross-coupling. The LZSM interference fringes are smooth and nearly symmetric, as clearly observed in Fig. 12 at high confinement ($\Omega = 500\omega$). This regime, referred to as the preserved symmetry regime, features a single perturbative avoided crossing dressed only by the lowest Bessel sideband. The spectrum displays highly linear, parallel, and stable energy bands, and multi-photon transitions are suppressed. The contour plot of transition probability reveals a clean, narrow resonant line centered at $\gamma_1 \approx k\omega/2$ (see Fig. 13, with $\Omega/2\pi$ = 60 GHz). Outside this line, $P \rightarrow 0.336$, demonstrating that strong spatial confinement effectively insulates the qubit from multi-photon LZSM interference. The geometric phase is strictly Abelian [$U(1)$], and the states remain widely separated by the large confinement gap, preserving an isolated, non-degenerate two-level manifold.
 
When the electrostatic potential relaxes ($\Omega \sim \alpha$, $\beta$), the spatial wavefunctions extend across the core-shell boundary, significantly increasing the effective coupling to the on-chip microwave antenna fields. As the gate voltage is reduced to lower $\hbar\Omega$ (weaker confinement), $\gamma_1$ increases quadratically, enabling traversal through multiple $k$-photon resonances within an experimentally accessible drive-amplitude range and resulting in concentric rings of true gap closings. Since $\gamma_2 = \alpha\beta/\Omega$ is odd under $\alpha \rightarrow -\alpha$ or $\beta \rightarrow -\beta$, while $\gamma_1$ and $\gamma_3$ are even, the resulting fringe pattern exhibits exact point (inversion) symmetry but lacks mirror symmetry. Once diabolical points emerge (as observed in Fig. 13 at low confinement, $\Omega = 0.8\omega$, with red circles indicating diabolical points or chiral crossings in multiphoton resonance branches), encircling these points generates genuinely chiral LZSM interferograms.
  
The contour plane of quasi-energies and transition probability evolves into a complex, nonlinear network of overlapping anticrossings (see Fig. 13 in the low-confinement regime, with $\Omega/2\pi$ = 8 GHz). The bands display a fractal-like or highly oscillatory structure determined by the intersections of zeros of multi-variable Bessel functions. This gate-voltage-driven emergence of nested gap closings, with chirality defined by the $\alpha$, $\beta$ cross-coupling and phase $\phi$, constitutes the confinement-induced topological Landau-Zener transition. Each diabolical point serves as a natural site for a non-Abelian holonomic gate in the GaAs/AlGaAs core-shell nanowire platform \cite{Lauhon, Funk, Morkotter}.

The contour map of $P_k$ reveals an intricate LZSM interference pattern, with regions of maximum transition probability alternating rapidly with regions of zero probability. Notably, the contour displays distinct nodes where the transition probability vanishes even under large drive amplitudes; these nodes correspond to the coordinates for Coherent Destruction of Tunneling (CDT) \cite{Grossman}. Consequently, the low-confinement regime functions as the experimental gateway to topological protection. As $\gamma_2$ and $\gamma_3$ attain high values, significant shifts cause different Floquet bands to intersect. At the precise coordinate nodes where CDT closes the effective gap ($\delta_k^{\text{eff}} \sim 0$), synthetic degeneracies arise. Electrostatically modulating the phase $\phi$ along a closed path around these degeneracy nodes induces non-commuting matrix rotations. Therefore, the low-confinement regime transforms the spatial drive into a non-Abelian synthetic gauge field, facilitating the implementation of holonomic quantum gates that are robust against local phase perturbations.
  
Since $\Delta\varepsilon/\omega$ and $P_k$ depend solely on the ratios ($\alpha/\omega$, $\beta/\omega$, $\delta/\omega$, $\Omega/\omega$), fixing $\omega$ at the upper end of the microwave range does not alter the topology of the quasi-energy surface; it only rescales the absolute energy and field axes. The nested $k$-photon resonance rings, diabolical points at their intersections, and the point-symmetric or chiral fringe pattern remain mathematically identical to those in the original $\omega$-normalized map, now expressed in GHz and, via the stated mapping, in $mT$. This demonstrates that the topological structure depends on the drive-amplitude-to-confinement ratio rather than the absolute drive frequency, consistent with the general high-frequency scaling theory of periodically driven systems \cite{Bukov}.

This connects the abstract topology defined by ($\alpha/\omega$, $\beta/\omega$) to material degrees of freedom: the conduction band offset $\Delta{E_c}$ at the GaAs/AlGaAs interface establishes the existence of a confined channel; the gate-tunable lateral parabola superimposed on this channel physically realizes $\Omega$ in the Hamiltonian; and the diabolical point indicated in panel (c) of Fig. 10, along with the encircling loop, represents the specific location where a Wilczek-Zee-type non-Abelian holonomy \cite{Wilczek, JPZanardi} could be implemented by cycling the gate voltage and microwave phase.

A crossing that persists as a true or near-true degeneracy under a periodic drive is the scenario addressed by driven-tunneling theory. Coherent destruction of tunneling (\cite{Grossman}; reviewed in \cite{Grifoni}) corresponds to the vanishing $\Delta\varepsilon$ observed at the ring intersections. It is this vanishing gap, rather than abstract topological arguments, that justifies interpreting a loop around it as a holonomic gate instead of an ordinary dynamical rotation.
  
Since $\Delta\varepsilon/\omega$ and $P_k$ depend solely on the ratios ($\alpha/\omega$, $\beta/\omega$, $\delta/\omega$, $\Omega/\omega$), fixing $\omega$ at the upper end of the microwave range does not alter the topology of the quasi-energy surface; it only rescales the absolute energy and field axes. The nested $k$-photon resonance rings, diabolical points at their intersections, and the point-symmetric or chiral fringe pattern remain mathematically identical to those in the original $\omega$-normalized map, now expressed in GHz and, via the stated mapping, in $mT$. This demonstrates that the topological structure depends on the drive-amplitude-to-confinement ratio rather than the absolute drive frequency, consistent with the general high-frequency scaling theory of periodically driven systems \cite{Bukov}.

This connects the abstract topology defined by ($\alpha/\omega$, $\beta/\omega$) to material degrees of freedom: the conduction band offset $\Delta{E_c}$ at the GaAs/AlGaAs interface establishes the existence of a confined channel; the gate-tunable lateral parabola superimposed on this channel physically realizes $\Omega$ in the Hamiltonian; and the diabolical point indicated in panel (c) of Fig. 10, along with the encircling loop, represents the specific location where a Wilczek-Zee-type non-Abelian holonomy \cite{Wilczek, JPZanardi} could be implemented by cycling the gate voltage and microwave phase.

A crossing that persists as a true or near-true degeneracy under a periodic drive is the scenario addressed by driven-tunneling theory. Coherent destruction of tunneling (\cite{Grossman}; reviewed in \cite{Grifoni}) corresponds to the vanishing $\Delta\varepsilon$ observed at the ring intersections. It is this vanishing gap, rather than abstract topological arguments, that justifies interpreting a loop around it as a holonomic gate instead of an ordinary dynamical rotation.

\subsubsection{Gate time}
\noindent 

Near an isolated $k$-photon resonance ($\gamma_1 = k\omega/2$), the two dressed levels exhibit Rabi-type oscillations at the Floquet gap frequency. Consequently, the time required to perform a full population-inverting ($\pi$-pulse) single-qubit gate is given by
\begin{equation}
t_{Gate} = \frac{\pi\hbar}{\Delta\varepsilon^k(\beta, \omega)},
\end{equation}
Here, $\Delta\varepsilon^k$ and $\delta^{k}_{\text{eff}}$ are as defined above.

At resonance, $\Delta\varepsilon^k = \delta^{k}_{\text{eff}}$, so $t_{Gate}$ is determined solely by the Bessel-suppressed multiphoton coupling. Off resonance, $\Delta\varepsilon$ is dominated by detuning and becomes large, which formally yields a short $\Delta\varepsilon$ but results in rapid, uncontrolled phase accumulation rather than a well-defined gate. The physically relevant regime is therefore the resonance-locked minimum of $\delta^{k}_{\text{eff}}$ and its vicinity. The exact gate time, $t_{Gate} = \frac{\pi\hbar}{\Delta\varepsilon^k}$, is plotted, reducing to the resonance expression and interpolating appropriately off resonance. The gate time $t_{Gate}$ is not monotonic with respect to the drive frequency $\omega$ in Fig. 15; it exhibits resonance-locked spikes (reaching up to approximately 16 ns for the low-confinement, $\beta/2\pi = 1.2$ GHz case, peaking near $\omega/2\pi = 0.44$ GHz), separated by fast-gate windows. The minimum values are around 0.45 to 0.6 ns near the upper end of the swept frequency range ($\omega/2\pi = 5$ GHz) for low confinement, and similarly around $1 ns$ for high confinement. Only 0.2\% (low confinement) and 0.1\% (high confinement) of the parameter grid exceed $t_{Gate} = 20 ns$, indicating that the divergences are sharp, localized features rather than broad regions \cite{Grossman}.
                                           
\begin{figure}
\begin{center}
	\includegraphics[width=14cm]{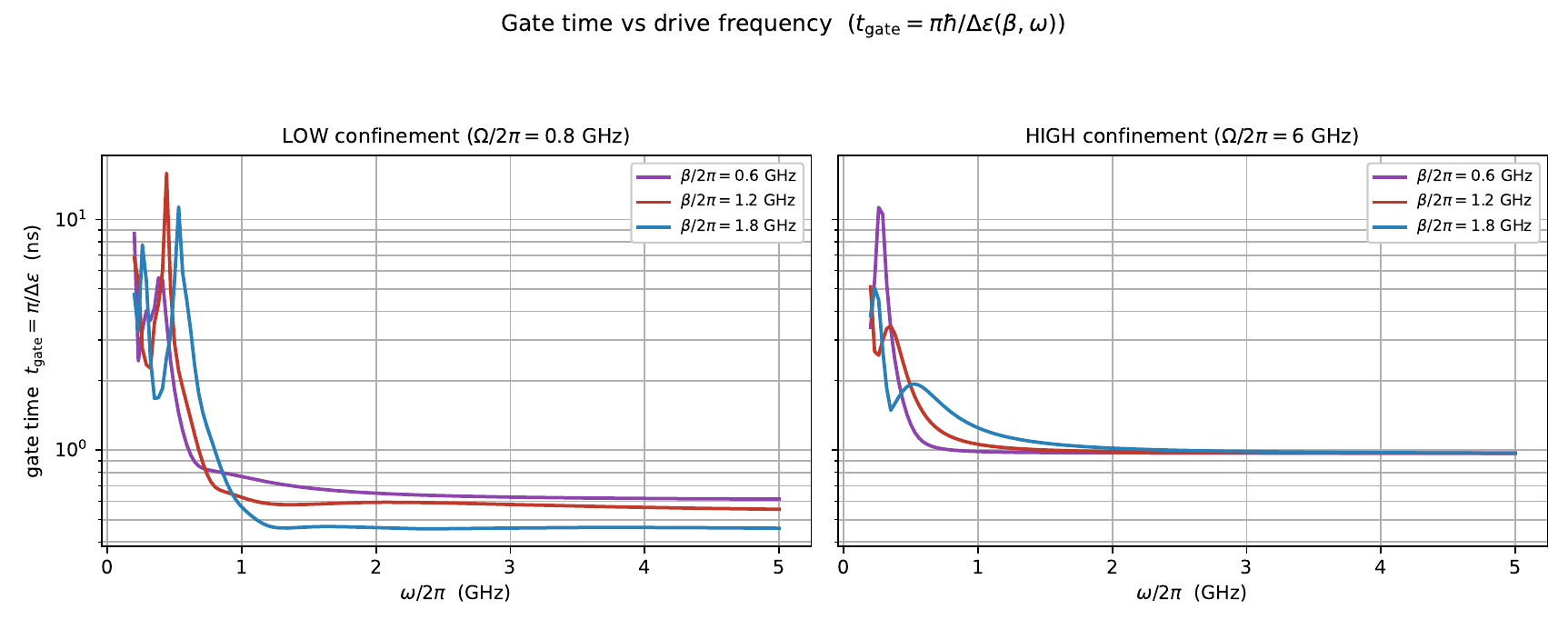}
\caption{(Color online)  The Gate-time vs drive frequency at low ($\Omega/2\pi$ = 0.8GHz) and high confinement regime ($\Omega/2\pi$ =6 GHz) for different value of homogenous driving amplitude $\beta/2\pi$ = \{0.8GHz, 1.2GHz,  1.8GHz\} at the fixed parameters  $\delta/2\pi$ = 0.4GHz, $\alpha/2\pi$ =1.2GHz  $\phi = \pi/3$.}
\end{center}
\end{figure}

\section{Discussion and conclusion}
\noindent

Theoretical analysis demonstrates that a spin qubit in a parabolic quantum wire, when driven by a bichromatic field, can be exactly reduced to a phase-modulated effective field in the RWA. This framework aligns with the principles of Floquet-LZSM interferometry, characterized by a secular rotation at rate $\gamma_1$ accompanied by Bessel-function sidebands determined by $\gamma_2/\omega$ and $\gamma_3/\omega$. Since all these parameters scale as $1/\Omega$, the confinement frequency $\Omega$ serves as a single control parameter, enabling continuous interpolation between a perturbative, single-resonance regime at high $\Omega$ and a genuinely multiphoton, multi-crossing regime at low $\Omega$, corresponding to the ``multi-crossing" structure introduced earlier.
 
Numerical results established three independent cuts of parameter space: ($\alpha$, $\beta$) at fixed $\omega$, and ($\beta$, $\omega$) at fixed $\alpha$. The exact Floquet monodromy calculation consistently reproduced the same qualitative transition.

$\bullet$ Under high confinement, $\gamma_1$ remains close to its bare value, with at most one resonance line entering any realistic sweep window. The quasi-energy gap remains strictly bounded away from zero, and the LZSM fringe pattern is smooth and only mildly chiral.

$\bullet$ Under low confinement, $\gamma_1$ traverses multiple half-integer resonance thresholds, resulting in nested rings or fans of near-gap-closing (diabolical) points in the initial $\alpha$, $\beta$ map (see Fig 12). In this regime, 148 grid points achieved $\Delta\varepsilon \leq 0.01$, with $\omega$ approaching zero in the high-confinement map. In the ($\beta$, $\omega$) map (see Fig 14), a similar asymmetry was observed, with approximately seven resonance orders visible rather than only one.

In each cut, the pattern exhibits exact inversion symmetry but broken mirror symmetry, as $\gamma_2 \propto \alpha\beta$ is odd under the inversion of either amplitude, whereas $\gamma_1$ and $\gamma_3$ are even. This provides a concrete and quantitative definition of "chiral interference": it is not merely a qualitative descriptor but rather a measurable, nonzero antisymmetric component that accounts for approximately 15\% of the total gap range in each map. This component becomes topologically significant only when true diabolical points are present and can be encircled. Collectively, these features constitute a confinement-induced topological Landau-Zener transition. The curve confinement $\Omega$ does not simply rescale the spectrum; instead, it enables or suppresses the existence of protected level crossings. Each crossing serves as a natural site for a Wilczek-Zee non-Abelian holonomy under cyclic transport in ($\alpha$, $\beta$, $\phi$) or ($\beta$, $\omega$, $\phi$) space.

$\bullet$  Analysis of Fig. 15 for the merit $t_{Gate}$ indicates that the two regimes are operationally complementary, rather than one being inherently superior. Off-resonance, both regimes support sub-nanosecond dynamical $\pi$-pulses. On resonance, near a diabolical point, $t_{Gate}$ diverges; this behavior is not indicative of failure but rather signifies that a slow, adiabatic, geometric (holonomic) protocol is the appropriate control strategy in a parabolic quantum wire, as opposed to a fast dynamical approach. Lower confinement increases the number of available holonomic stations per unit drive amplitude or frequency range.

$\bullet$ In the context of the proposed GaAs/AlGaAs high-mobility core-shell nanowire, each theoretical control parameter directly corresponds to an established lithographic or microwave capability. Surface split-gates enable continuous, electrostatic tuning of $\hbar\Omega$. On-chip coplanar antennas provide the two quadrature drive tones at $\omega/2\pi$, with amplitudes generating $mT$-scale fields. The relative phase $\phi$ is controlled electronically. No additional fabrication steps are necessary beyond those already used in high-mobility core-shell nanowire growth and gate patterning. The requirements of the theory, including tunable parabolic confinement, biharmonic microwave delivery, and phase control, are all achievable with current technology. This alignment of theoretical demands with existing capabilities positions the proposal as a near-term, rather than speculative, advancement. The following scenario is specified:

(i) The mapping between dimensionless and physical units presented here serves as a convenient normalization. In the source literature, models of this type typically reference a GaAs device operating at $\hbar\Omega$ = 1-10 $meV$, which, in absolute terms, is situated deep within the high-confinement (perturbative) regime relative to $mT$-scale drive amplitudes. Accessing the multi-crossing regime in practice requires either substantially larger drive amplitudes or operation at the lower end of the confinement window. This trade-off should be evaluated in relation to the achievable field or voltage ranges of a specific fabricated device.

(ii) The present calculation considers an isolated, closed two-level system. In practical implementations, real gate times and transition probabilities will be constrained by charge noise, phonon-mediated relaxation, and residual valley or orbital coupling, which are not included in this analysis.

``Non-Abelian holonomy" strictly requires a genuinely degenerate, rather than a quasi-degenerate, subspace to be transported adiabatically. The numerical analysis presented here identifies the locations of near-degeneracies and demonstrates how their density depends on confinement, which is a necessary precondition. However, a complete non-Abelian gate construction would require an explicit multi-sideband degenerate-perturbation-theory treatment, which has only been qualitatively outlined above. In this model, the confinement frequency of a gated GaAs/AlGaAs core-shell nanowire serves as a definitive topological control parameter. Reducing this frequency does not merely alter the qubit speed; it activates a family of protected quasi-energy crossings that serve both as resources for non-Abelian holonomic gates and as bottlenecks for fast dynamical gates. This duality provides the platform with an intrinsic, voltage-tunable choice between two paradigms of quantum control. The topological transition can be accessed either by tuning ($\alpha$, $\beta$) at fixed $\omega$ or by tuning $\omega$ at fixed $\alpha$, $\beta$, with confinement determining the density of diabolical points in either scenario. For gate design, this implies that fast-gate windows (sub-nanosecond, off-resonance-adjacent) are optimal for dynamical (Rabi-type) single-qubit gates, while the divergent $t_{Gate}$ resonance spikes correspond to diabolical points where a slow, adiabatic loop in ($\beta$, $\omega$, $\phi$) yields a controlled non-Abelian holonomy. Experimental realization of these effects is feasible with current nanofabrication and microwave ($\omega/2\pi$ = 10 GHz) control technologies, and the proposed interferometric protocol offers a direct measurement approach. Collectively, these findings position quantum-wire materials as a versatile, scalable, and robust platform for topological quantum control, quantum simulation, and fault-tolerant quantum information processing \cite{Zanardi, Jones, LuMing, Solinas, Katabarwa}. Principal results include the identification of a confinement-induced topological transition, the emergence of non-Abelian geometric phases, and the observation of Floquet-Bloch oscillations within parameter space. The study is further advanced by presenting a concrete experimental blueprint using GeSi semiconductor heterostructures, establishing a pathway to scalable multi-qubit entanglement via synthetic gauge fields, and introducing a framework for quantifying intrinsic decoherence resilience through Dynamical Symmetry. Additionally, the engineering of a Floquet Quantum Wire Interferometer for metrology and the potential integration with machine learning for optimal control are proposed.

\begin{acknowledgements}
\noindent
This work was supported by the Organization for Women in Science for the Developing World ({\color{blue}OWSD}) and the Swedish International Development Cooperation Agency ({\color{blue}SIDA}). 
\end{acknowledgements}

\newpage

\appendix
\counterwithin{equation}{section}  
\renewcommand{\theequation}{A\arabic{equation}}  
\section{}

The transformed Hamiltonian $\bar{H}(t)$ follows from the general relation below.
\begin{equation} 
\bar{H}(t) = V_0^{\dag}H(t)V_0(t) - i\hbar{V_0^{\dag}\dot{V}_0(t)}.
\end{equation}

\subsection{Computation of $- i\hbar{V_0^{\dag}\dot{V}_0(t)}$} 
\noindent

Given $V_0(t) = \exp{(-\frac{i}{2\hbar}\phi(t)\hat{\sigma}_z)}$, the time derivative is calculated as follows:
\begin{equation}
\dot{V}_0(t) = -\frac{i\dot{\phi}(t)}{2\hbar}\hat{\sigma}_zV_0(t).
\end{equation}
Thus,
\begin{equation}
- i\hbar{V_0^{\dag}\dot{V}_0(t)} = -\frac{\dot{\phi}(t)}{2}\hat{\sigma}_z.
\end{equation}

\subsection{Transformation of the diagonal component $-\frac{h(t)}{2}\hat{\sigma}_z$}
\noindent

The operator $\hat{\sigma}_z$ commutes with any function of $\hat{\sigma}_z$, including $V_0(t)$.
\begin{equation}
V_0^{\dag}\Big(-\frac{h(t)}{2}\hat{\sigma}_z\Big)V_0(t) - i\hbar{V_0^{\dag}\dot{V}_0(t} = -\frac{h(t)}{2}\hat{\sigma}_z.
\end{equation}
The derivative $\dot{\phi}(t)$ is obtained by differentiating Eq. (6) term by term.

\begin{equation}
  \begin{split}
\dot{\phi}(t) & =   \gamma_1 - \frac{\gamma_2}{\hbar\omega}\cdot{\hbar\omega}\cdot\sin(\omega{t})-\frac{\gamma_3}{2\hbar\omega}\cdot{2\hbar\omega}\cdot\cos(2\omega{t} + \theta) \\
        &= \gamma_1 - {\gamma_2}\sin(\omega{t}) - {\gamma_3}\cos(2\omega{t} + \theta) \\
        &= h(t).
\end{split}
\end{equation}
This result is notable because, by construction, $\dot{\phi}(t) = h(t)$. Substituting this relation into Eq. (A3),
\begin{equation}
- i\hbar{V_0^{\dag}\dot{V}_0(t)} = -\frac{h(t)}{2}\hat{\sigma}_z
\end{equation}

\subsection{ Assembly and exact cancellation of the diagonal term}
\noindent

By summing all contributions from Eqs. (A8), and (A9),
\begin{equation}
  \begin{split}
\bar{H}(t) & =   \underbrace{-\frac{h(t)}{2}\hat{\sigma}_z}_{\text{from} V_0^{\dag}{H}V_0} + \underbrace{\Bigg(\frac{h(t)}{2}\hat{\sigma}_z\Bigg)}_{\text{ gauge term} - i\hbar{V_0^{\dag}\dot{V}_0(t)}}\cdot(-1)  - \frac{\Delta}{2}\big[\hat{\sigma}_x\cos{\phi(t)} + \hat{\sigma}_y\sin{\phi(t)}\big] \\
           &=   \underbrace{-\frac{h(t)}{2}\hat{\sigma}_z + \frac{h(t)}{2}\hat{\sigma}_z }_{= 0} - \frac{\Delta}{2}\big[\hat{\sigma}_x\cos{\phi(t)} + \hat{\sigma}_y\sin{\phi(t)}\big].             
\end{split}
\end{equation}
The terms involving ${h(t)}$ in $\hat{\sigma}_z$ cancel exactly. The transformation ${V}_0(t)$ is constructed to eliminate the time-dependent bias $h(t)$ from the Hamiltonian. The resulting Hamiltonian is equivalently expressed in matrix form as follows:
\begin{equation}
\bar{H}(t) = -\frac{\Delta}{2}\begin{pmatrix}
0&e^{-i\phi(t)}\\e^{+i\phi(t)}&0
\end{pmatrix}.
\end{equation}

\renewcommand{\theequation}{B\arabic{equation}}  
\section{}
\noindent

\subsection{Jacobi-Anger expansion applied to $\bar{H}(t)$}
\noindent

The fundamental Jacobi-Anger identities are as follows:

\begin{subequations}
\begin{align}
e^{iz\sin\theta} & = \sum_{n = -\infty}^{\infty}J_n(z)e^{i{n}\theta}, \\
e^{iz\cos\theta} & = \sum_{n = -\infty}^{\infty}i^nJ_n(z)e^{i{n}\theta}. 
\end{align}
\end{subequations}
Here, $J_n$ denotes the Bessel functions of the first kind.

\subsection{Expansion of $e^{\pm{i\phi{t}}}$}
\noindent

Eq. (6) in the text defines the phase function introduced by the canonical transformation as follows:
\begin{equation}
\phi(t) =   \gamma_1{t} - \frac{\gamma_2}{\hbar\omega}\cos(\omega{t}) - \frac{\gamma_3}{2\hbar\omega}\sin(2\omega{t} + \theta) + K(\gamma_2, \gamma_3).
\end{equation}
From this equation, $e^{\pm i\phi{t}}$ can be expanded by factorizing and separating the individual contributions:
\begin{equation}
e^{\pm{i\phi{t}}} =   e^{\pm{i\gamma_1{t}}}\cdot{e^{\pm{iK(\gamma_2, \gamma_3) }}}\cdot{e^{\pm{i\frac{\gamma_2}{\hbar\omega}\cdot\cos(\omega{t})}}}\cdot{e^{\pm{i\frac{\gamma_3}{2\hbar\omega}\sin(2\omega{t} + \theta)}}}.
\end{equation}
The Jacobi-Anger identity is utilized for both oscillatory factors as described in Eqs. (B1) as follows:
\begin{subequations}
\begin{align}
  e^{\pm{i}\frac{\gamma_2}{\hbar\omega}\cos(\omega{t})}             & = \sum_{n = -\infty}^{\infty}i^nJ_n\Big(\frac{\gamma_2}{\hbar\omega}\Big)e^{\pm{i}n\omega{t}}, \\
  e^{\pm{i}\frac{\gamma_3}{2\hbar\omega}\sin(2\omega{t} + \theta)}  & = \sum_{n = -\infty}^{\infty}J_m\Big(\frac{\gamma_3}{2\hbar\omega}\Big)e^{\pm{i}{m}(2\omega{t} + \theta)}. 
\end{align}
\end{subequations}
By combining Eqs. (B3), and (B4), the exponential $e^{\pm i\phi(t)}$ can be rewritten in the following compact form:
\begin{equation}
e^{\pm{i\phi{t}}} = e^{\pm{i}K(\gamma_2, \gamma_3) }\sum_{n, m = -\infty}^{\infty}i^nJ_n\Big(\frac{\gamma_2}{\hbar\omega}\Big){\cdot}J_m\Big(\frac{\gamma_3}{2\hbar\omega}\Big)\cdot{e^{\pm{i{m}\theta}}}e^{\pm{i}[\gamma_1 + (n - 2m)\omega]t}.
\end{equation}
Substituting Eq. (B5) into Eq. (A8) yields:
\begin{equation}
\bar{H}(t) = -\frac{\Delta}{2}\sum_{n, m = -\infty}^{\infty}i^nJ_n\Big(\frac{\gamma_2}{\hbar\omega}\Big){\cdot}J_m\Big(\frac{\gamma_3}{2\hbar\omega}\Big)\cdot{e^{\pm{i{m}\theta}}}\begin{pmatrix}
0&e^{-{i}K(\gamma_2, \gamma_3)}e^{+{i}[\gamma_1 + (n - 2m)\omega]t}\\e^{+{i}K(\gamma_2, \gamma_3) }e^{-{i}[\gamma_1 + (n - 2m)\omega]t}&0
\end{pmatrix}.
\end{equation}


\end{document}